\documentclass[a4paper,number,5p,twocolumn]{elsarticle}

\usepackage[english]{babel}
\usepackage[utf8]{inputenc}
\usepackage{graphicx}
\usepackage[usenames,dvipsnames]{xcolor}
\usepackage{bm}
\usepackage{array}
\usepackage{amsfonts}
\usepackage{amssymb}
\usepackage{amsmath}
\usepackage{epstopdf}
\usepackage{soul}
\usepackage[normalem]{ulem}
\usepackage{lineno}
\usepackage{setspace}
\usepackage{csquotes}
\usepackage{algorithm}
\usepackage{algpseudocode}
\usepackage[
    unicode,
    breaklinks=true,
    hypertexnames=false,
    colorlinks=false,
    citecolor=black,
    filecolor=black,
    linkcolor=black,
    urlcolor=black
]{hyperref}
\usepackage{natbib}
\usepackage{mathtools}
\usepackage{interval}
\usepackage{multirow}
\usepackage{paralist}
\usepackage{xifthen}
\usepackage{comment}

\newcommand{\Fref}[1]{Fig.~\ref{#1}}

\newcommand{\Sref}[1]{{Section~\ref{#1}}}

\newcommand{\etal}{~et~al.}

\newcommand{\romannumber}[1]{\uppercase\expandafter{\romannumeral #1\relax}}

\mathchardef\mhyphen="2D

\newcommand{\tens}[1]{\boldsymbol{#1}}					% A first-order tensor
\renewcommand{\hl}[1]{#1}

\DeclareGraphicsExtensions{.pdf,.eps,.png,.jpg}	

\soulregister\cite7
\soulregister\ref7
\soulregister\pageref7

\makeatletter
\providecommand{\doi}[1]{%
	\begingroup
	\let\bibinfo\@secondoftwo
	\urlstyle{rm}%
	\href{http://dx.doi.org/#1}{%
		doi:\discretionary{}{}{}%
		\nolinkurl{#1}%
	}%
	\endgroup
}
\makeatother

\begin{document}

% ----------------------------------------
% Front matter
% ----------------------------------------

\begin{frontmatter}

\title{%
Wang tiles enable combinatorial design and robot-assisted manufacturing of modular mechanical metamaterials}

\author[mech]{Martin Do\v{s}k\'{a}\v{r}}
\ead{martin.doskar@cvut.cz}
\author[mech]{Michael Somr}
\ead{michael.somr@fsv.cvut.cz}
\author[mech,exp]{Radim Hlůžek}
\ead{radim.hluzek@fsv.cvut.cz}
\author[exp]{Jan Havelka}
\ead{jan.havelka@fsv.cvut.cz}
\author[exp]{Jan Nov\'{a}k}
\ead{novakja@fsv.cvut.cz}
\author[mech]{Jan Zeman}
\ead{jan.zeman@cvut.cz}
\address[mech]{Department of Mechanics, Faculty of Civil Engineering, Czech Technical University in Prague, Th\'{a}kurova 7, \mbox{166 29 Prague 6}, Czech Republic}
\address[exp]{Experimental centre, Faculty of Civil Engineering, Czech Technical University in Prague, Th\'{a}kurova 7, \mbox{166 29 Prague 6}, Czech Republic}

\journal{arXiv.org}

\begin{abstract}
	In this paper, we introduce a novel design paradigm for modular architectured materials that allows for spatially nonuniform designs from a handful of building blocks, which can be robotically assembled for efficient and scalable production. 
	The traditional, design-limiting periodicity in material design is overcome by utilizing Wang tiles to achieve compatibility among building blocks.
	We illustrate our approach with the design and manufacturing of an L-shaped domain inspired by a scissor-like soft gripper, whose internal module distribution was optimized to achieve an extreme tilt of a tip of the gripper's jaw when the handle part was uniformly compressed. The geometry of individual modules was built on a 3$\times$3 grid of elliptical holes with varying semi-axes ratios and alternating orientations. 
	We optimized the distribution of the modules within the L-shaped domain using an enumeration approach combined with a factorial search strategy. 
	To address the challenge of seamless interface connections in modular manufacturing, we produced the final designs by casting silicone rubber into modular molds automatically assembled by a robotic arm. The predicted performance was validated experimentally using a custom-built, open-hardware test rig, Thymos, supplemented with digital image correlation measurements. 
	Our study demonstrates the potential for enhancing the mechanical performance of architectured materials by incorporating nonuniform modular designs and efficient robot-assisted manufacturing.
\end{abstract}


\end{frontmatter}

% ----------------------------------------
% Introduction
% ----------------------------------------
\section{Introduction}

Metamaterials are the prime manifestation of the structure-property relations in architectured materials. Their remarkable macroscopic response, often unprecedented in nature, stems from a purposely designed microstructural geometry rather than the bulk properties of individual components. Because of their exotic properties, metamaterials receive attention in many areas of physics, from electromagnetism~\cite{Watts_2012, Barroso_2020} to optics~\cite{Ren_2019, Manzoor_2021}, acoustics~\cite{Hedayati_2020, Zhou_2020}, or solid mechanics~\cite{ziemke_chirality_2019, Jenett_2020}, and find applications in medicine~\cite{Wang_2016, Chen_2018}, soft robotics~\cite{rafsanjani_programming_2019, Wen_2020}, solar power management~\cite{Chauhan_2020, Liu_2012}, and civil engineering~\cite{Achaoui_2016, Klett_2017}.
\begin{figure*}[!ht]
	\centering
	\begin{tabular}{ccc}
		\scalebox{0.9}{%% Creator: Inkscape 1.1.1 (3bf5ae0d25, 2021-09-20), www.inkscape.org
%% Accompanies image file 'module_v2.pdf' (pdf, eps, ps)
\begingroup%
  \makeatletter%
  \providecommand\color[2][]{%
    \errmessage{(Inkscape) Color is used for the text in Inkscape, but the package 'color.sty' is not loaded}%
    \renewcommand\color[2][]{}%
  }%
  \providecommand\transparent[1]{%
    \errmessage{(Inkscape) Transparency is used (non-zero) for the text in Inkscape, but the package 'transparent.sty' is not loaded}%
    \renewcommand\transparent[1]{}%
  }%
  \providecommand\rotatebox[2]{#2}%
  \newcommand*\fsize{\dimexpr\f@size pt\relax}%
  \newcommand*\lineheight[1]{\fontsize{\fsize}{#1\fsize}\selectfont}%
  \ifx\svgwidth\undefined%
    \setlength{\unitlength}{133.51516964bp}%
    \ifx\svgscale\undefined%
      \relax%
    \else%
      \setlength{\unitlength}{\unitlength * \real{\svgscale}}%
    \fi%
  \else%
    \setlength{\unitlength}{\svgwidth}%
  \fi%
  \global\let\svgwidth\undefined%
  \global\let\svgscale\undefined%
  \makeatother%
  \begin{picture}(1,0.99999992)%
    \lineheight{1}%
    \setlength\tabcolsep{0pt}%
    \put(0,0){\includegraphics[width=\unitlength,page=1]{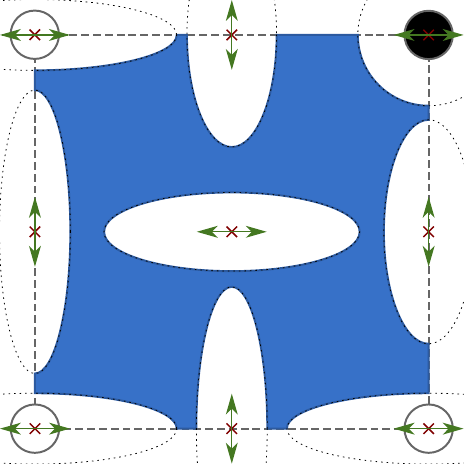}}%
    \put(0.07584478,-0.02570062){\makebox(0,0)[t]{\lineheight{1.25}\smash{\begin{tabular}[t]{c}$\kappa_{1} = 4.0$\end{tabular}}}}%
    \put(0.925467,-0.02570062){\makebox(0,0)[t]{\lineheight{1.25}\smash{\begin{tabular}[t]{c}$\kappa_{2} = 4.0$\end{tabular}}}}%
    \put(0.07584478,0.99665521){\makebox(0,0)[t]{\lineheight{1.25}\smash{\begin{tabular}[t]{c}$\kappa_{3} = 4.0$\end{tabular}}}}%
    \put(0.925467,0.99665521){\makebox(0,0)[t]{\lineheight{1.25}\smash{\begin{tabular}[t]{c}$\kappa_{4} = 1.0$\end{tabular}}}}%
  \end{picture}%
\endgroup%
} &
		\includegraphics[height=4cm]{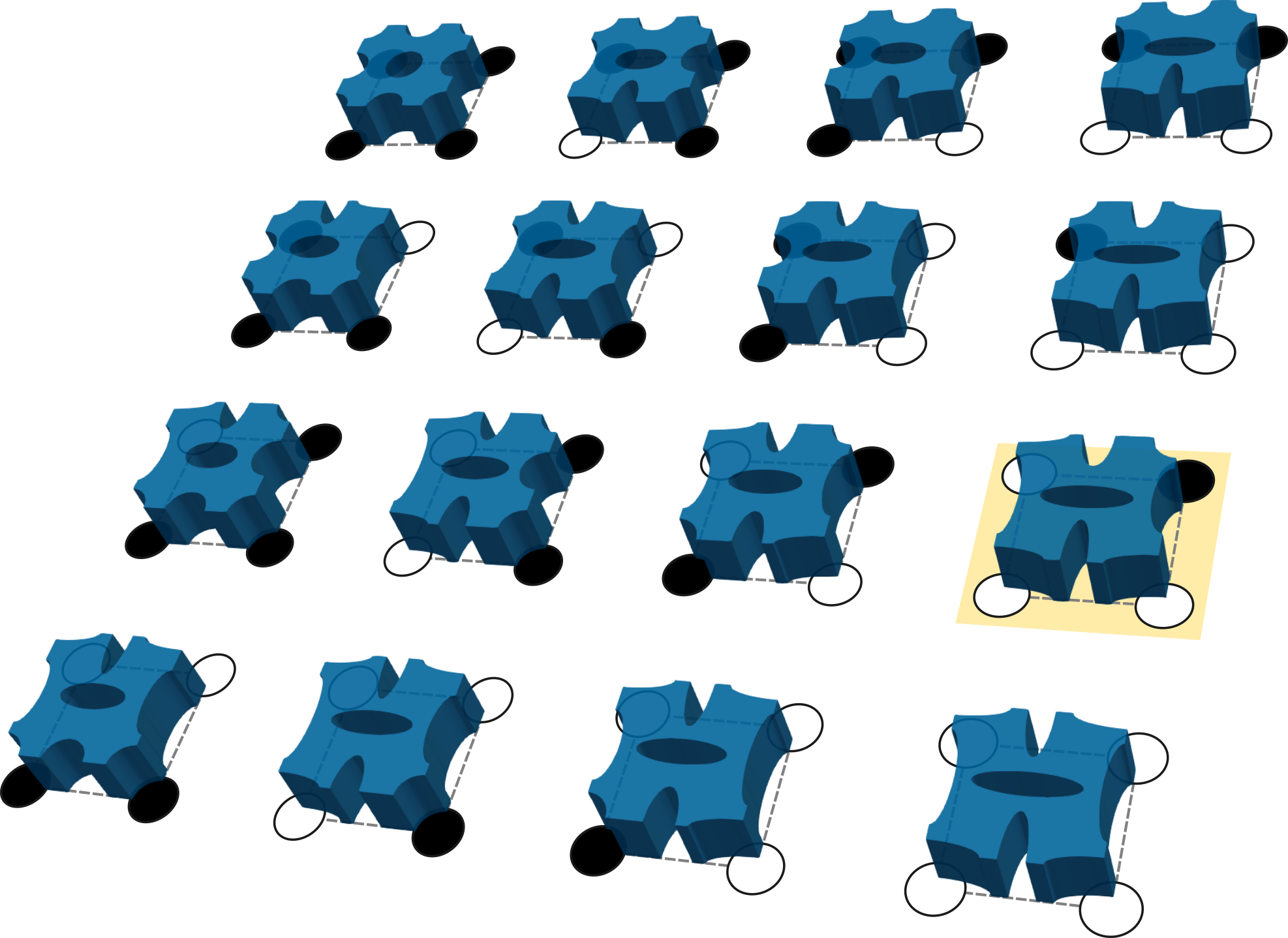} & 
		\scalebox{1.0}{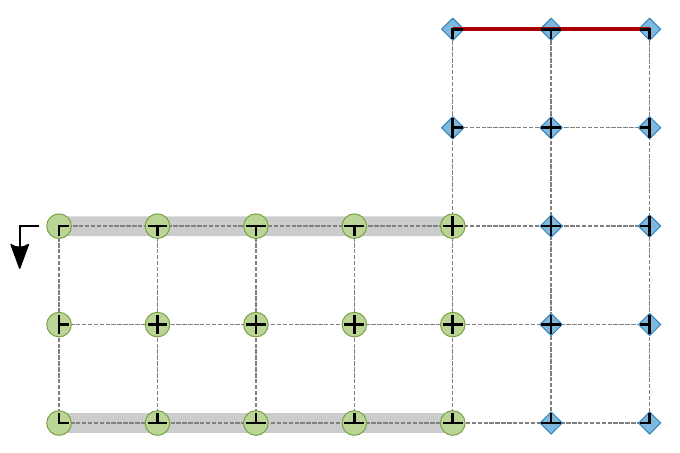} \\
		(a) & (b) & (c)
	\end{tabular}
	\caption{(a) Blueprint of a module design comprising a 3$\times$3 grid of elliptical holes with fixed orientation (denoted with green arrows), fixed area $A = 0.18^{2} \pi$, and varying ratio $\kappa$ of major to minor semi-axes' lengths given by bi-linear interpolation of vertex-related values $\kappa_{i}$ pertinent to vertex codes ($\kappa = 1.0$ or $4.0$ for black and white codes, respectively); (b) Visualization of the resulting set (the highlighted module corresponds to the design shown in (a)). (c) Illustration of the assembly-level problem. The domain is supported along the $\textsf{AB}$ edge, prescribed displacement $\widehat{\tens{v}}$ with magnitude corresponding to $5\%$ vertical strain acts on the edge $\textsf{DE}$. The objective of optimizing the distribution of black and white codes in green circular and blue rhombic nodes is to achieve extreme (maximum and minimum) inclination angle $\phi$ of the top part $\textsf{FG}$ (measured from the linear term in the second-order polynomial approximation of vertical displacement of the part; see~\Fref{fig:composed_histograms} for an illustration of $\phi$).The distinction between green and blue nodes follows from the factorial search described in~\mbox{\Sref{sec:design}}, with vertex codes in green nodes optimized first.}
	\label{fig:layout_scheme}
\end{figure*}

In this work, we focus on mechanical metamaterials~\cite{Buckmann_2014, Paulose_2015, Coulais_2015}. These can exhibit negative Poisson's ratio (auxeticity)~\cite{Lakes_1987, Mousanezhad_2015}, high strength-to-density ratio~\cite{Bauer_2016}, self-recovery of shock absorbers~\cite{Frenzel_2016}, multistability~\cite{Nicolaou_2012, Florijn_2014, Waitukaitis_2015}, and programmability~\cite{Silverberg_2014, He2020, jin_kirigamiinspired_2020}. To push the application limits further, recent designs have utilized hierarchical approaches~\cite{Lakes_1993, Li_2017, Berwind_2018, gu_bionspired_2018a}, drawn inspiration from ancient Japanese arts of paper folding (origami)~\cite{Wickeler_2020, Dieleman2020} or cutting (kirigami)~\cite{Tang_2016, jin_kirigamiinspired_2020}, and employed machine learning~\cite{Xue_2020, Wu2020, Ma_2020} or inverse analysis~\cite{Matlack_2018, Oliveri2020} algorithms. 

Although the structural complexity of mechanical metamaterials is vast, most designs consist of a periodically repeated cell resulting in spatially homogeneous properties~\cite{Grima_2006, mullin_pattern_2007, Schaedler_2011}. Yet entirely new functionalities can arise when considering non-periodic architectures. Coulais et al.'s~\cite{Coulais2016} combinatorial approach to the design of non-periodic metamaterials with spatially textured properties enabled anisotropic deformation of building blocks forming a cubic arrangement deforming into a predefined two-dimensional pixelated surface texture. A strategy of matching a bandgap evinced by a non-periodic metamaterial assembled from three different building blocks and its periodic counterpart was presented by D'Alessandro et al.~\cite{D_Alessandro_2020}. Reid et al.~\cite{Reid_2018} demonstrated that disordered networks and their pruning can exhibit a tunable auxetic response. Analytical methods for designing origami closed-loop units (and predicting their folding response) that can be used as building blocks for modular construction of foldable structures with desired performance were presented by Mousanezhad et al.~\cite{Mousanezhad2017}. Another origami-based strategy enables designing complex and pluripotent (capable of folding into multiple shapes) crease patterns with the help of 140 distinct foldable motifs---jigsaw puzzle pieces~\cite{Dieleman2020}. Jin et al.~\cite{jin_kirigamiinspired_2020} applied kirigami principles to design programmable inflatables that can mimic target shapes. Capabilities of machine learning were also exploited to design a bioinspired hierarchical composite~\cite{gu_bionspired_2018a} or modular metamaterial with specified properties~\cite{Wu2020} and to create a framework accelerating the characterization and pattern generation of a non-uniform cellular material~\cite{Ma_2020}.

Our contribution elaborates a new class of combinatorial mechanical metamaterials stemming from the concept of Wang tiling. Foundations of this concept date back to Hao Wang~\cite{wang_proving_1961}, who recast the satisfiability problem for first-order logic into the generation of aperiodic coverings of the plane formed by a finite number of tile types, which must satisfy edge-matching constraints in analogy to assembling jigsaw puzzle pieces. As such, Wang tiling provides a unified description of periodic and non-periodic patterns while employing a handful of building blocks with predefined edge connectivity, with subsequent applications in passive DNA self-assembly~\cite{winfree_design_1998,yan_dna-templated_2003,tikhomirov_fractal_2017} (with a potential to be extended to centimeter scales~\cite{jilek2021,Jilek_2022}), computer graphics~\cite{cohen_wang_2003,sibley_wang_2004,kopf_recursive_2006,liu_fabricable_2022}, or modeling random heterogeneous materials~\cite{doskar_aperiodic_2014,doskar_wang_2018,doskar_jigsaw_2016,doskar_level-set_2020,doskar_microstructure-informed_2021}.

By leveraging the inherent modularity of Wang tiles, we present a full workflow of the parameterization, optimal design, and robot-assisted fabrication of soft modular metamaterials of locally tunable auxeticity. To this goal, in~\Sref{sec:design} we introduce the adopted Wang tile set and its geometry, along with a simple computational procedure to determine the tile assemblies that deliver the target mechanical performance. \Sref{sec:manufacturing} deals with sample production, with an emphasis on automating the entire process via prefabrication. Subsequently, the computational model is validated against experimental measurements in~\Sref{sec:experiments}, and the study concludes with several extensions outlined in~\Sref{sec:conclusions}.

% ----------------------------------------
% Design
% ----------------------------------------
\section{In-silico design}
\label{sec:design}
Adopting the formalism of Wang tiles, the design procedure comprises two levels~\cite{tyburec_modular-topology_2021,tyburec_modular-topology_2022}: 
(a) the design of individual modules and (b) their distribution within a given upper-scale domain. 
The modular design thus avoids the requirement of separation of scales typical for multi-scale setups, in which any material point in a domain carries a (potentially distinct) microstructural geometry, and it represents a compromise between designs based on one optimized cell, e.g.~\cite{wu_topology_2021}, and traditional single-scale topology optimization designs with fully resolved structural details, e.g.,~\cite{aage_giga-voxel_2017}.

In this work, we employed the vertex-based definition of Wang tiles~\cite{lagae_alternative_2006} with codes attributed to tile vertices instead of tile edges, as this variant allows for a control of the corner-connected neighbors and thus remedies the main drawback of the edge-based variant~\cite{cohen_wang_2003}.
We demonstrated our framework on a design comprising 16 modules stemming from the vertex-based Wang set over two vertex codes; see~\Fref{fig:layout_scheme}b.
Topologically, each module is a square domain with a regular 3$\times$3 grid of elliptical holes (or their parts) with a constant area $A$ and alternating orientation. 
The final shape of the holes---and consequently the whole module geometry---was governed by the ratios $\kappa$ of the major to minor semiaxes lengths, which were obtained by a bilinear interpolation of values $\kappa_{i}$ pertinent to the individual vertex codes $i = 1\dots4$; see~\Fref{fig:layout_scheme}a.
In our setup, we chose $\kappa_{i} \in\left\{ 1.0, 4.0 \right\}$ for black and white vertex codes, respectively, and $A = 0.18^{2} \pi$ for a unit-size module. 
Consequently, the interpolation allowed for a smooth transition from auxetic regions with elliptical holes~\cite{mullin_pattern_2007,bertoldi_mechanics_2008} to pattern-transforming regions with circular holes~\cite{mullin_pattern_2007,van_bree_newton_2020}.

\begin{figure*}[ht]
	\centering
%	\scalebox{0.85}{\input{composed_results.pdf_tex}} 
	\includegraphics[width=0.95\textwidth]{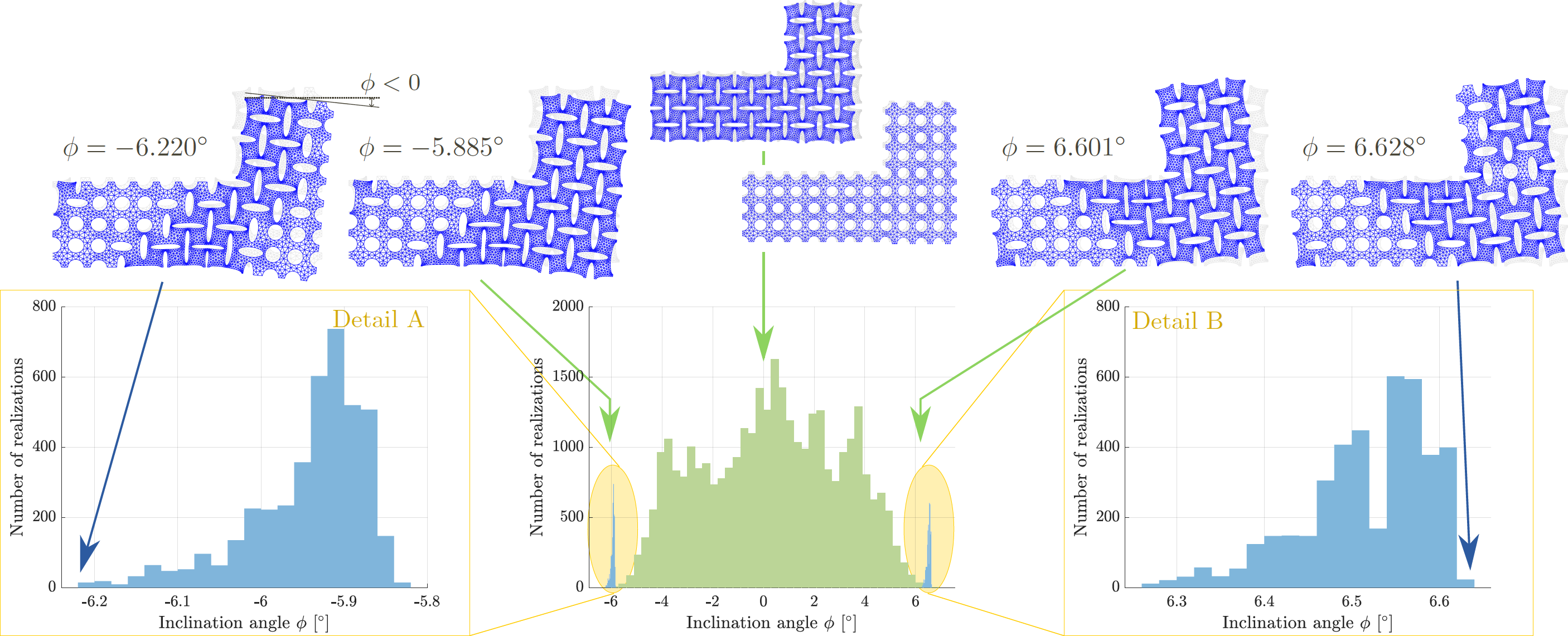}
	\caption{Results of the factorial exploration of the combinatorial design space. The central histogram shows in green the inclination angles $\phi$ achievable by changing the codes in green circular points in~\Fref{fig:layout_scheme}c with the remaining codes being fixed as white (i.e., $\kappa = 4.0$). The highlighted blue parts of the histogram (with zoomed-in details A and B on the left- and right-hand sides of the figure) depict the distribution of $\phi$ arising from the second exploration phase of the factorial approach, in which the codes at green circular points are fixed at values identified in the first phase and the remaining codes (blue square points in~\Fref{fig:layout_scheme}c) are optimized. The upper row illustrates the resulting design pertinent to extremes from all histograms (complemented with the results of the two uniform designs).}
	\label{fig:composed_histograms}
\end{figure*}

At the macroscopic assembly level, we considered a 6~module wide/4~module tall L-shape domain shown in~\Fref{fig:layout_scheme}c.
The domain was loaded by prescribed vertical displacement $\widehat{\tens{v}}$ inducing $5\%$ vertical strain acting on the top edge $\textsf{DE}$, and was vertically supported along the part $\textsf{AB}$ of the bottom edge (to recall the notation of edges, see~\Fref{fig:layout_scheme}c). To avoid rigid body translations in simulations, the left-most node on $\textsf{AB}$ edge was also fixed in the horizontal direction.
This choice was inspired by soft-robotics applications of architectures materials for grippers, e.g.\mbox{~\cite{yang_buckling_2015,sinatra_ultragentle_2019}}. In particular, the macroscopic domain resembled a half of a scissor-like gripper actuated by uniform compression of its handle part. Consequently,
as an objective of the assembly design, we tried to reach the extreme inclination angles $\phi$ (both minimum and maximum\hl{, for the sake of completeness}) of the upper part $\textsf{FG}$ by varying the placement of modules within the domain. The target extreme angles were computed from a linear term in the least-square second-order polynomial approximation of the vertical displacements of all nodes along the edge $\textsf{FG}$.

%\hl{Without required modularity and the predefined design of modules, topology or shape optimization approaches---applied to the design of architectured materials, e.g., in~\cite{wang_isogeometric_2017,jeong_shape_2019,medina_nonlinear_2023} and \cite{sigmund_tailoring_1995,wang_3d_2021}, respectively--could have been employed. 
\hl{If neither modularity was mandated nor the design of modules was predetermined, topology~\cite{sigmund_tailoring_1995,osanov_topology_2016,wang_3d_2021} or shape~\cite{wang_isogeometric_2017,jeong_shape_2019,medina_nonlinear_2023} optimization methods could have been considered.
However, since we started with a fixed module design, the optimization problem outlined above was inherently combinatorial. An approximate solution might be obtained using stochastic meta-heuristic algorithms, such as the Genetic Algorithm~\cite{Holland_1992,matous_applying_2000,chen_maximizing_2023}---already successfully used in our previous work on bi-level optimization of truss-like Wang tiles~\cite{tyburec_modular-topology_2021}, or simulated annealing~\cite{Oliveri2020}.
Alternatively, outputs of the shape optimization might have served as a guidance for module distribution; similarly to the Free Material optimization we employed in a more general setup for simultaneous optimization of continuum modules and their assembly in~\cite{tyburec_modular-topology_2022}. 
Nonetheless, for the sake of simplicity, we kept the optimization part as simple as brute-force evaluation in this letter to focus on the whole workflow, and refer an interested reader to, e.g., \cite{tyburec_modular-topology_2021,tyburec_modular-topology_2022} for our dedicated optimization strategies for modular structures and mechanisms.}
% leaving dedicated optimization strategies which we have been developing simultaneously to separate publications, e.g., above-mentioned references~\cite{tyburec_modular-topology_2021,tyburec_modular-topology_2022}.}

Even though the domain was small and thus contained a low number of module positions, the full combinatorial design space encompassed $2^{27}$ (more than $134$ million) assemblies.
\hl{To circumvent the necessity of exploring the entire design space, we employed a factorial search strategy, following our observations that the distribution of vertex codes in the \mbox{$\textsf{ABED}$} block---as illustrated in~\mbox{\Fref{fig:layout_scheme}c}--exerts the most pronounced impact on the response. Consequently, we initially conducted an exhaustive evaluation of all attainable combinations involving variations in the vertex codes relevant to the lower section (represented as green, circular points \mbox{in~\Fref{fig:layout_scheme}c)}. After identifying extreme cases within this subset, we fixed the identified design and performed a subsequent search, this time altering the vertex codes in the remaining areas of the domain} (i.e., in blue, rhombic points in~\Fref{fig:layout_scheme}c).
Consequently, we performed only $2^{15} + 2 \times 2^{12} = 40,960$ simulations in total.

\begin{figure*}[ht]
	\centering
	\scalebox{0.8}{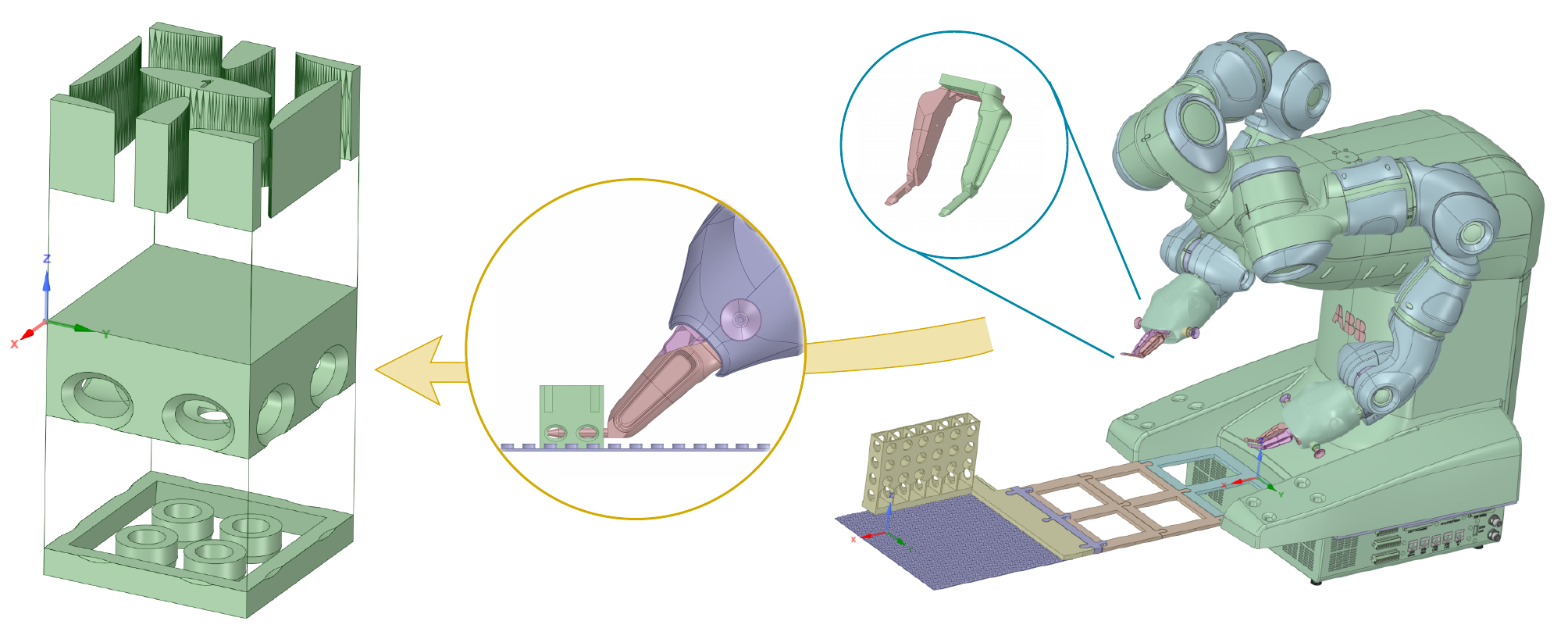} 
	\caption{(a) An expanded view of a modular mold composed of three parts (the top section with the actual module geometry imprint, the middle section with holes for a fork-lift pair of grippers, the bottom section with cylindrical studs to attach to LEGO\textsuperscript{\textregistered} baseplates). (b) Scheme of the robotic assembly setup using ABB robot IRB 1400 YuMi equipped with a custom 3D-printed pair of grippers (Inset 2), distancers (1), silo-like feeder (2), and LEGO\textsuperscript{\textregistered} baseplate (3) to facilitate the assembly and attachment of the modules into a final mold (see Inset 1).}
	\label{fig:robotic_assembly}
\end{figure*}

All simulations were performed with an in-house code available at a Zenodo repository~\cite{doskar_2023_zenodo}, which extends the finite element toolbox for finite strain calculations accompanying the work of van Bree et al.~\cite{van_bree_newton_2020}. Each module was discretized with quadratic triangular elements using the T3D mesh generator~\cite{Rypl_2022_t3d}. 
The behavior of the bulk material was modeled with a compressible neo-Hookean hyper-elastic constitutive law with an additional term added by Bertoldi\etal{}~\cite{bertoldi_mechanics_2008} to better fit the response of a silicone rubber at larger strains; see \textit{Supplementary materials} for the particular form of a strain-energy density function. 
Parameters of the constitutive law were taken as reported by Bertoldi et al.~\cite{bertoldi_mechanics_2008} and used, e.g., in~\cite{rokos_extended_2020,van_bree_newton_2020}.
A contact mechanism was not incorporated in the numerical model as both simulation and validation experiments did not exhibit closing of the voids.

Results of the brute-force evaluation are shown in~\Fref{fig:composed_histograms}. In uniform designs, i.e., assembly plans in which all vertex values are the same, there was no inclination of the top edge $\textsf{FG}$ as the whole domain laterally either shrank (for the auxetic arrangement of elliptical holes with $\kappa = 4.0$) or expanded (for circular holes with $\kappa = 1.0$).
Performing an enumeration search for the green points revealed that both extreme angles are achieved by formation of a triangular auxetic region in either \textsf{ABE} ($\phi < 0$) or \textsf{BED} ($\phi > 0$) part of the domain.
The subsequent explorations through the remaining blue points only led to minor improvements over previous designs (with more pronounced improvement in the negative angle compared to the positive one); see the inset figures on left- and right-hand sides of~\Fref{fig:composed_histograms}.

\hl{Admittedly, there is no guarantee that such a procedure results in a global optimum. Yet, the only minor improvements achieved in the second part of the factorial search support our assumption of the dominant role of vertex codes within the $\textsf{ABED}$ region.
Furthermore, subsequent factorial searches conducted (i) with the alternate initial setup beyond the $\textsf{ABED}$ region and (ii) with an expanded initial region encompassing all 21 vertex locations in the bottom part of the domain---please refer to Appendix B in the \textit{Supplementary materials} for comprehensive specifics---yielded analogous extreme angles as those listed in~\mbox{\Fref{fig:composed_histograms}}, thereby affirming the reasonable quality of the optimized designs.}

\begin{figure*}[]
	\centering 
%	\scalebox{0.7}{\input{real_images.pdf_tex}} 
	\includegraphics[width=0.95\textwidth]{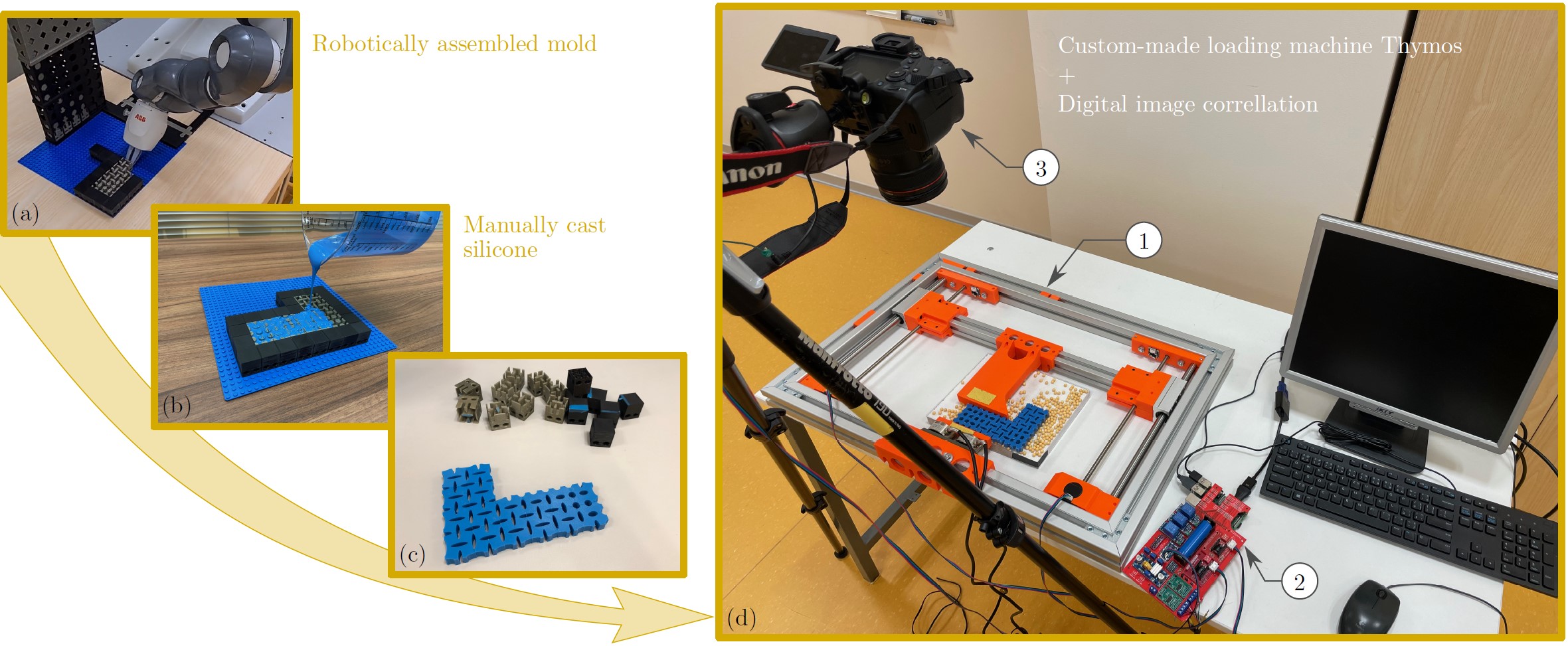}
	\caption{Snapshots of the manufacturing process: (a) modules assembled by the YuMi robot, (b) silicone rubber cast into the assembled mold, (c) final specimen after demolding, and (d) overview of the loading setup (\textcircled{1} - custom-made Thymos loading frame, \textcircled{2} - Thymos control unit, \textcircled{3} - digital camera for digital image correlation measurements). The robotic assembly and the process of silicone casting and demolding were also captured in videos available on \mbox{\href{https://youtu.be/ko5P2HqI4WI}{youtu.be/ko5P2HqI4WI}} and \mbox{\href{https://youtu.be/rlhP6sQ_kVE}{youtu.be/rlhP6sQ\_kVE}}.}
	\label{fig:real_photos}
\end{figure*}

\begin{figure*}[]
	\centering
	\scalebox{0.30}{%% Creator: Inkscape 1.2.1 (9c6d41e410, 2022-07-14), www.inkscape.org
%% Accompanies image file 'vertex_based_validation_composite_results_v2.pdf' (pdf, eps, ps)
\begingroup%
  \makeatletter%
  \providecommand\color[2][]{%
    \errmessage{(Inkscape) Color is used for the text in Inkscape, but the package 'color.sty' is not loaded}%
    \renewcommand\color[2][]{}%
  }%
  \providecommand\transparent[1]{%
    \errmessage{(Inkscape) Transparency is used (non-zero) for the text in Inkscape, but the package 'transparent.sty' is not loaded}%
    \renewcommand\transparent[1]{}%
  }%
  \providecommand\rotatebox[2]{#2}%
  \newcommand*\fsize{\dimexpr\f@size pt\relax}%
  \newcommand*\lineheight[1]{\fontsize{\fsize}{#1\fsize}\selectfont}%
  \ifx\svgwidth\undefined%
    \setlength{\unitlength}{1786.9802496bp}%
    \ifx\svgscale\undefined%
      \relax%
    \else%
      \setlength{\unitlength}{\unitlength * \real{\svgscale}}%
    \fi%
  \else%
    \setlength{\unitlength}{\svgwidth}%
  \fi%
  \global\let\svgwidth\undefined%
  \global\let\svgscale\undefined%
  \makeatother%
  \begin{picture}(1,0.60736469)%
    \lineheight{1}%
    \setlength\tabcolsep{0pt}%
    \put(0,0){\includegraphics[width=\unitlength,page=1]{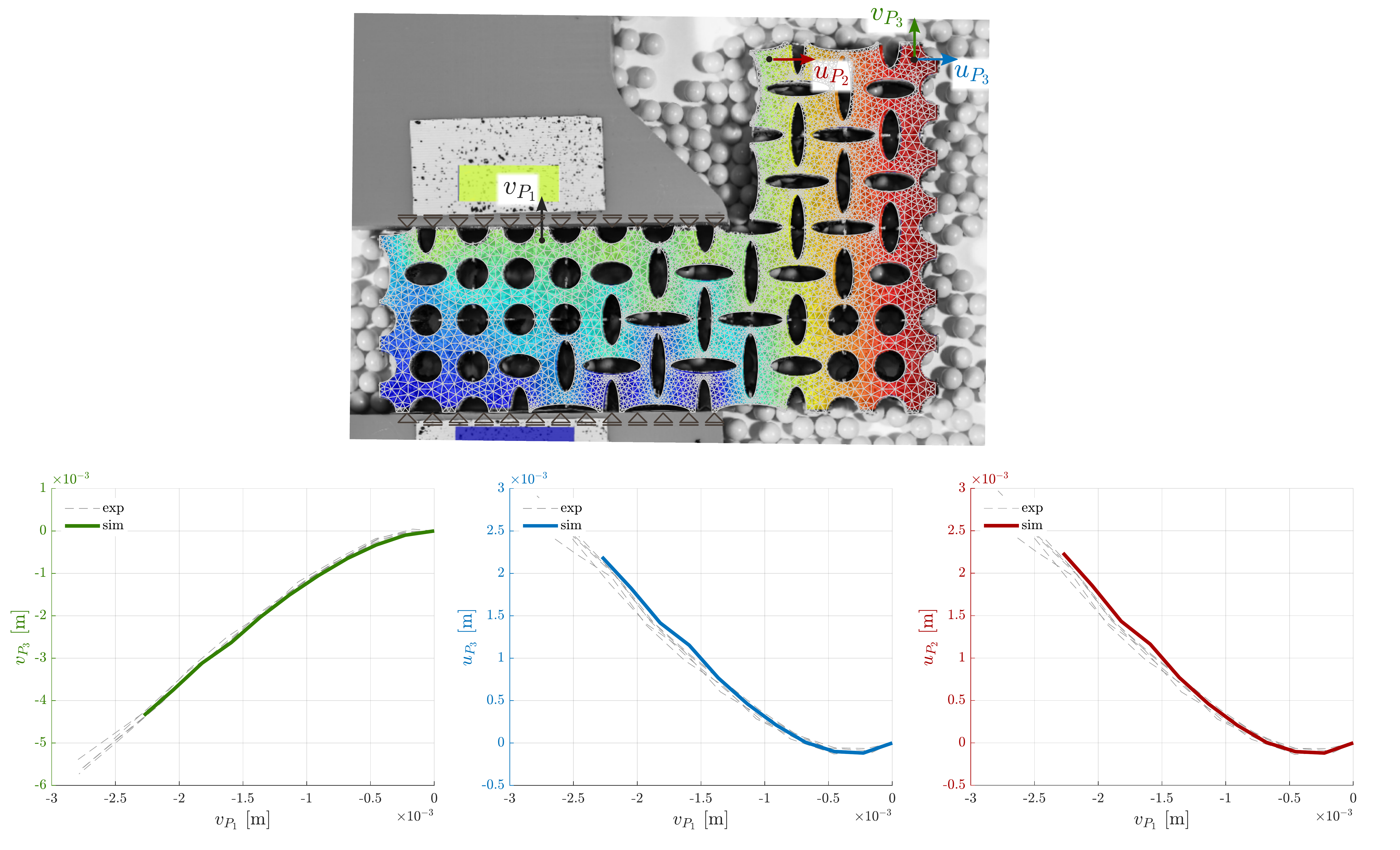}}%
    \put(0.20690971,0.28112188){\makebox(0,0)[lt]{\lineheight{1.25}\smash{\begin{tabular}[t]{l}\Huge{(a)}\end{tabular}}}}%
    \put(0.65647257,0.00420453){\makebox(0,0)[lt]{\lineheight{1.25}\smash{\begin{tabular}[t]{l}\Huge{(d)}\end{tabular}}}}%
    \put(-0.00022952,0.00529969){\makebox(0,0)[lt]{\lineheight{1.25}\smash{\begin{tabular}[t]{l}\Huge{(b)}\end{tabular}}}}%
    \put(0.32717454,0.00437894){\makebox(0,0)[lt]{\lineheight{1.25}\smash{\begin{tabular}[t]{l}\Huge{(c)}\end{tabular}}}}%
    \put(0,0){\includegraphics[width=\unitlength,page=2]{vertex_based_validation_composite_results_v2.pdf}}%
  \end{picture}%
\endgroup%
} 
	\caption{Comparison of computed and measured displacements: (a) Composite image showing experimental results from digital image correlation (plotted in an undeformed configuration) with the overlaid Finite Element discretization and position of selected points (marked with black filled circles), whose measured displacements, parameterized with the to prescribed vertical displacement, were compared against computed values in (b)--(d). Red filled squares in the scheme mark centers of ligaments where horizontal displacements were measured using DIC to incorporate the frictional behavior into a numerical model.}
	\label{fig:results}
\end{figure*}

% ----------------------------------------
% Robot-assisted manufacturing
% ----------------------------------------
\section{Robot-assisted manufacturing}
\label{sec:manufacturing}

Although the design phase already benefited from modularity (e.g., it reduced the design space~\cite{tyburec_modular-topology_2021,tyburec_modular-topology_2022} and accelerated calculations by employing domain-decomposition techniques~\cite{doskar_wang_2018}), 
modularity---together with prefabrication---appeals mainly to the production phase, as it allows overcoming the limits of manufacturing processes related to the interplay between manufacturing precision and size constraints.
In our previous work~\cite{nezerka_jigsaw_2018}, which illustrated this potential already with a single rotatable module with jigsaw puzzle locks, we highlighted the importance of interface bonding: to mimic a monolithic assembly, the puzzle locks needed to be very tight, which then hindered easy manipulation.
Here, we pursued a complementary approach to manufacturing in which the resulting specimen is cast into modular, robotically assembled molds.

The molds for individual modules were composed of three horizontally stacked parts, illustrated in \Fref{fig:robotic_assembly}a.
The top part comprised the negative form of the corresponding module design, whose geometry was automatically generated from the design parameters introduced in Section~\ref{sec:design}.
The middle part contained a pair of elliptical holes on each side, which were designed to facilitate robotic manipulation by a custom-made gripper (shown in the inset in \Fref{fig:robotic_assembly}c). Each hole led through the whole block and was conically widened at both ends to compensate for inaccuracies during robotic manipulation.
Finally, the bottom part of each mold block followed the standard geometry of a 3$\times$3 stud wide LEGO\textsuperscript{\textregistered} brick such that the block could be easily attached to a LEGO baseplate; further details are provided in the \textit{Supplementary materials}.

In addition to the regular module blocks described above, we also manufactured three types of auxiliary blocks which served as a formwork around the assembled mold (one for straight edge, one for inner and one for outer corner). 
The molds were manufactured from a standard PLA (polylactic acid) material using a conventional Prusa i3 MK3S+ hobby 3D printer.

For the automated assembly of the printed molds in the final cast, we utilized a dual-arm  ABB YuMi (IRB 14000 YuMi) collaborative robot.
We equipped the robotic arms with a 3D-printed pair of grippers inspired by a forklift to manipulate the molds.
The robotic arms picked the molds from an extensible silo-like feeder and place them at a 32$\times$32 studs LEGO baseplate, whose position relative to the feeder and the robot itself was ensured with planar distancers, see Figs.~\ref{fig:robotic_assembly}b and \ref{fig:real_photos}a. 
The order and distribution of the molds in the feeder were predefined in a RobotStudio script controlling the arms a priori; our implementation did not deal with automatic recognition of the mold types in the feeder. Again, more details are provided in the \textit{Supplementary materials}. Moreover, a video of the automated robotic assembly of the modular molds is available on \href{https://youtu.be/ko5P2HqI4WI}{YouTube}.

Lastly, Havel Composites ZA 22 two-component silicone was cast into the assembled molds, see~\Fref{fig:real_photos}b, and, after curing, the final sample was manually demolded and cleared of flash molding defects spanning the cross-modular holes as shown in~\Fref{fig:real_photos}c and in the second video available on \href{https://youtu.be/rlhP6sQ_kVE}{YouTube}.

% ----------------------------------------
% Experiments
% ----------------------------------------
\section{Validation of in-silico design}
\label{sec:experiments}

The manufactured samples were tested using a custom-built Thymos\footnote{\href{http://thymos.cz/}{http://thymos.cz/}} loading machine (shown in~\Fref{fig:real_photos}d) with displacements measured optically using Digital Image Correlation (DIC); see \textit{Supplementary materials} for additional information. 
%The tests were performed in a horizontal setup to suppress out-of-plane buckling.
\hl{All tests were performed in a horizontal setup to mitigate the influence of self-weight on in-plane deflection and to suppress out-of-plane displacement caused by inaccurate in-plane alignment of the sample in the loading machine.}
Since our preliminary experiments demonstrated a significant influence of friction on the obtained data, we placed the specimens on loose plastic airsoft pellets, which played the role of planar ball bearings.
However, the same adjustment could not easily be performed for the machine's (vertical) loading and supporting plates, and we observed a stick-slip behavior of material ligaments in contact with the plates.
To account for this frictional behavior, we measured a horizontal displacement of the center of each material ligament (highlighted with red filled squares in the scheme in~\Fref{fig:results}a) and consequently imposed these values as additional Dirichlet boundary conditions to the numerical model.
The comparison of the calculated against measured displacements (related to the prescribed vertical displacement) for selected points of the design minimizing the tilt angle $\phi$ is depicted in~\Fref{fig:results} and demonstrates an excellent match between the computational design and the measured data; see \textit{Supplementary materials} for analogous results for the design yielding the maximal $\phi$.

% ----------------------------------------
% Conclusions and outlook
% ----------------------------------------
\section{Conclusions and discussion}
\label{sec:conclusions}

This work introduced a design paradigm for modular architectured materials along with a full workflow covering design, optimization and robot-assisted fabrication of a modular mechanical metamaterial.
With a fixed module geometry inspired by classical auxetic and pattern-transforming metamaterials, the optimized assemblies resulting in the minimum and maximum inclination angles $\phi$ of the L-shape domain's top part $\textsf{FG}$ (recall~\Fref{fig:layout_scheme}c) were obtained by combinatorial evaluation of a numerical, finite-strain model. 
The designs identified by numerical simulations were turned into reality with robot-assisted manufacturing. To avoid difficulties related to seamless binding of modules, we exploited modularity in an earlier manufacturing stage: specimens were produced by casting silicone rubber into a modular mold assembled by a robotic arm. 
Response of the specimens was then tested with an open-hardware loading machine, Thymos, which enabled testing in a horizontal setup. The influence of uncertainty caused by stick-slip friction between a silicone specimen and loading plates was incorporated in the numerical model via prescribed evolution of horizontal displacement of contact ligaments, obtained from DIC. 
The outcomes of experimental validation confirmed that the mechanical response of a modular assembly can be accurately designed and tuned using a full combinatorial design---accelerated by factorial search---and numerical analysis. The accuracy of the proposed framework is demonstrated by the comparison of simulations and experiments, showing perfect agreement in the initial stages of loading. The maximum discrepancy in the final stages did not exceed 10\%, prior to self-contact not accounted for in our numerical model. 

Since the presented design of individual modules was driven mainly by engineering intuition, our current endeavors are focused on (i) modifying the standard topology optimization approaches for the purpose of modular design and (ii) developing optimizers suitable for the discrete problem of assembly layouts.
We have already made the first steps in this direction with a bilevel optimization scheme that combines optimal elastic design for truss modules with a meta-heuristics approach addressing the combinatorial problem~\cite{tyburec_modular-topology_2021}, which we recently extended to continuum models combined with mechanics-guided clustering heuristics~\cite{tyburec_modular-topology_2022} and manufacturing constraints~\cite{tyburec_modular-topology_2023}.

% ----------------------------------------
% CRediT
% ----------------------------------------
\section*{CRediT}
	\noindent
	MD - Conceptualization, Methodology, Software, Validation, Formal analysis, Data Curation, Visualization, Writing - Original Draft;
	MS - Investigation, Resources, Writing - Original Draft;
	RH - Investigation, Writing - Review \& Editing;
	JH - Methodology, Software, Resources, Writing - Review \& Editing;	
	JN - Conceptualization, Resources, Writing - Review \& Editing;
	JZ - Conceptualization, Supervision, Funding acquisition, Project administration, Writing - Review \& Editing.

% ----------------------------------------
% Acknowledgement
% ----------------------------------------
%\clearpage
\section*{Acknowledgment}	
The authors thank the Intelligent and Mobile Robotics (IMR) group at the Czech Institute of Informatics, Robotics and Cybernetics, CTU in Prague for granting us an access to the dual-arm ABB IRB 14000 YuMi collaborative robot.
We also thank Stephanie Krueger for a critical review and proof-reading of the initial versions of this manuscript\hl{ and anonymous reviewers for their helpful suggestions and comments}.
This research was funded by the Czech Science Foundation, project No. 19-26143X.

% ----------------------------------------
% Bibliography
% ----------------------------------------

\end{document}

\begin{frontmatter}
	
	\title{Wang tiles enable combinatorial design and robot-assisted manufacturing of modular mechanical metamaterials\\Supplementary materials}
	
	\author[mech]{Martin Do\v{s}k\'{a}\v{r}}
	\ead{martin.doskar@cvut.cz}
	\author[mech]{Michael Somr}
	\ead{michael.somr@fsv.cvut.cz}
	\author[mech,exp]{Radim Hlůžek}
	\ead{radim.hluzek@fsv.cvut.cz}
	\author[exp]{Jan Havelka}
	\ead{jan.havelka@fsv.cvut.cz}
	\author[exp]{Jan Nov\'{a}k}
	\ead{novakja@fsv.cvut.cz}
	\author[mech]{Jan Zeman}
	\ead{jan.zeman@cvut.cz}
	\address[mech]{Department of Mechanics, Faculty of Civil Engineering, Czech Technical University in Prague, Th\'{a}kurova 7, \mbox{166 29 Prague 6}, Czech Republic}
	\address[exp]{Experimental centre, Faculty of Civil Engineering, Czech Technical University in Prague, Th\'{a}kurova 7, \mbox{166 29 Prague 6}, Czech Republic}
	
	\journal{arXiv.org}
	
\end{frontmatter}

% ----------------------------------------
% Supplementary materials
% ----------------------------------------
\appendix
\section{Numerical model}

All numerical simulations were performed using an in-house codebase developed for finite element analyses in Matlab.
In particular, we adopted a finite-strain model with a hyper-elastic constitutive law for a silicone rubber material under the plane strain conditions.
The Bertoldi-Boyce~\cite{bertoldi_mechanics_2008} type of strain energy density function $W(\tenss{C})$ in the form
\begin{equation}
	W(\tenss{C}) = c_{1} \left( I_{1} - 3 \right) + c_{2} \left( I_{1} - 3 \right)^{2} - 2 c_{1} \log J + \frac{K}{2} \left( J - 1 \right)^{2}
\end{equation}
was used, 
with invariants $I_{1}$ and $J$ of the right Cauchy-Green tensor $\tenss{C}$ defined as $I_{1} = \tr \tenss{C}$ and $J = \sqrt{\det \tenss{C}}$; see discussion in~\ref{sec:validation_mat} for the particular choice of coefficient values.

Each module was discretized individually with quadratic triangular elements using the T3D meshing tool~\cite{Rypl_2022_t3d}, which allows for maintaining topologically and geometrically compatible meshes across selected boundaries of different modules.
Consequently, a finite element discretization of a modular aggregate can be readily assembled from the module discretizations, sparing us the need to remesh each modular assembly.

The numerical model was loaded with a prescribed vertical displacement $\hat{\tens{v}}$ imposed at the upper edge of the horizontal arm of the domain, i.e., edge $\textsf{DE}$ in~ Fig.~1c in the main manuscript. All vertical displacements of nodes at the bottom edge $\textsf{DE}$ were restricted and, in order to avoid rigid-body horizontal translation, the left-most node at $\textsf{DE}$ was also fixed horizontally.
The numerical solver was based on the standard Newton algorithm for non-linear optimization with a direct sparse solver in each iteration. Two load increments were used to reach the final displacement $\hat{\tens{v}}$. \hl{One simulation took approximately $2.5~\text{s}$ on a workstation with Intel\textsuperscript{\textregistered{}} Core\texttrademark{} i9-9900K and $128~\text{GB}$ RAM, running MATLAB 2022a in serial.}

\section{Factorial search}

\begin{figure}
	\centering
	\begin{tabular}{cc}
		\includegraphics[width=0.35\textwidth]{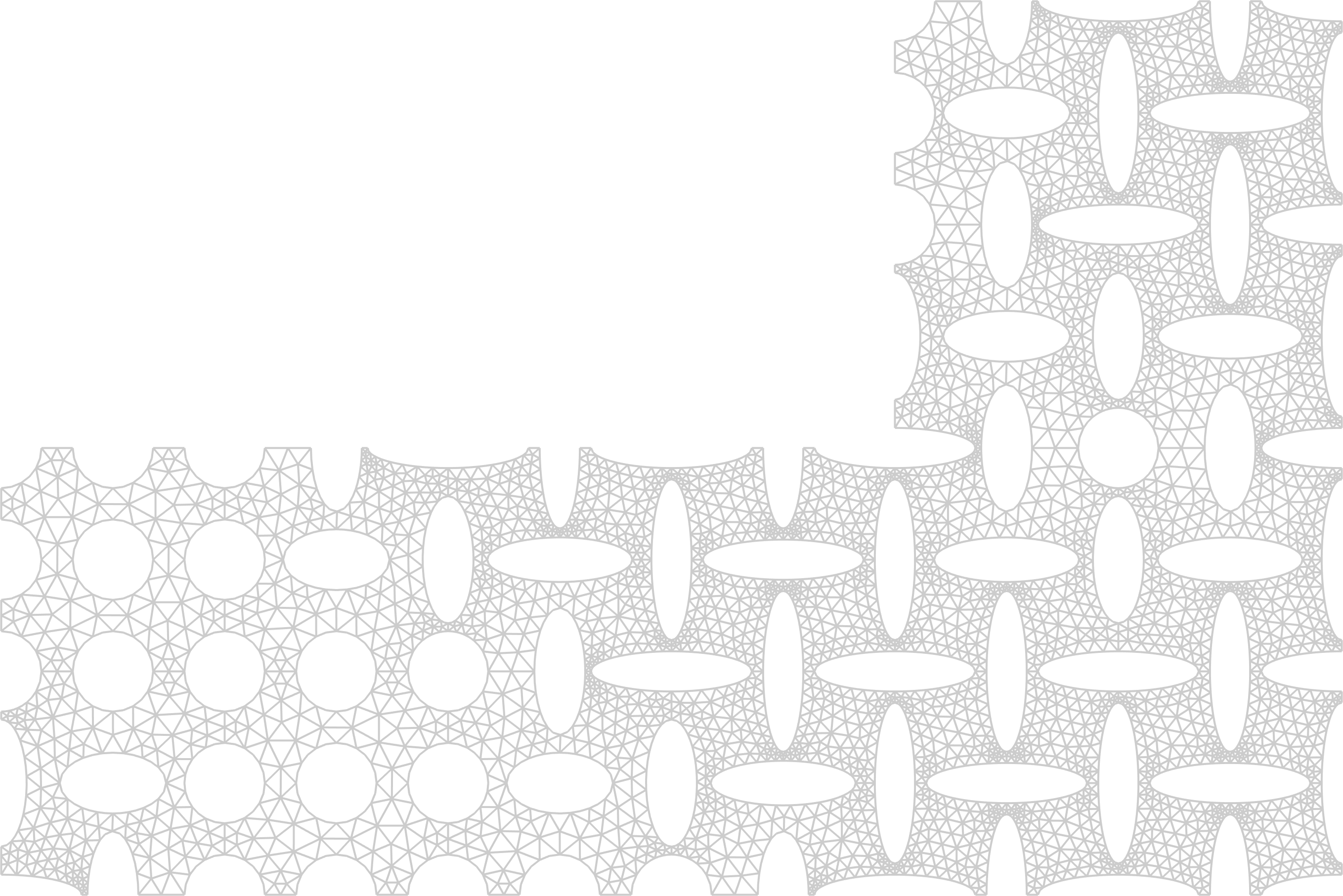} &
		\includegraphics[width=0.35\textwidth]{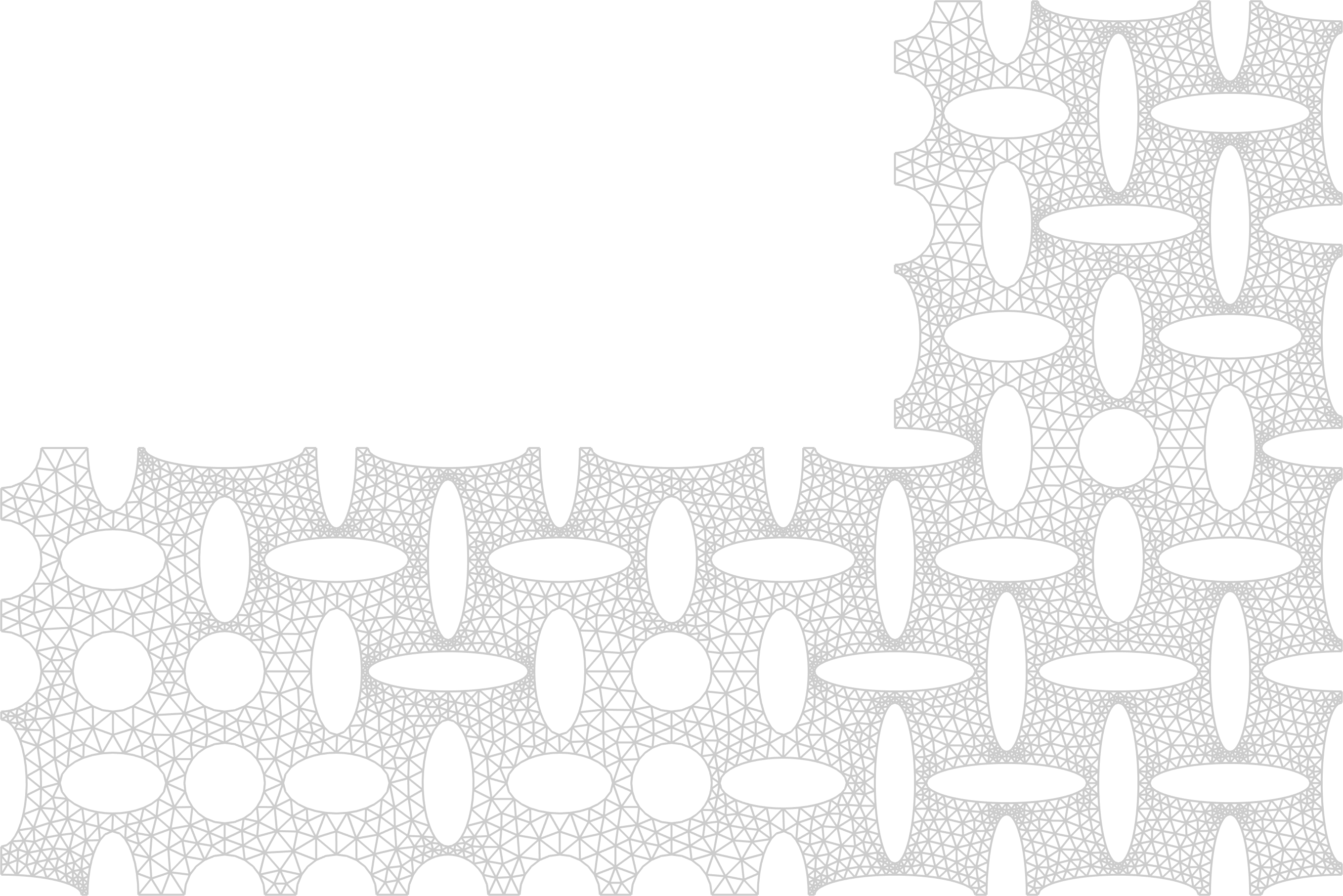} \\
		$\phi = 6.628^{\circ}$ & $\phi = 6.624^{\circ}$ \\
		\includegraphics[width=0.35\textwidth]{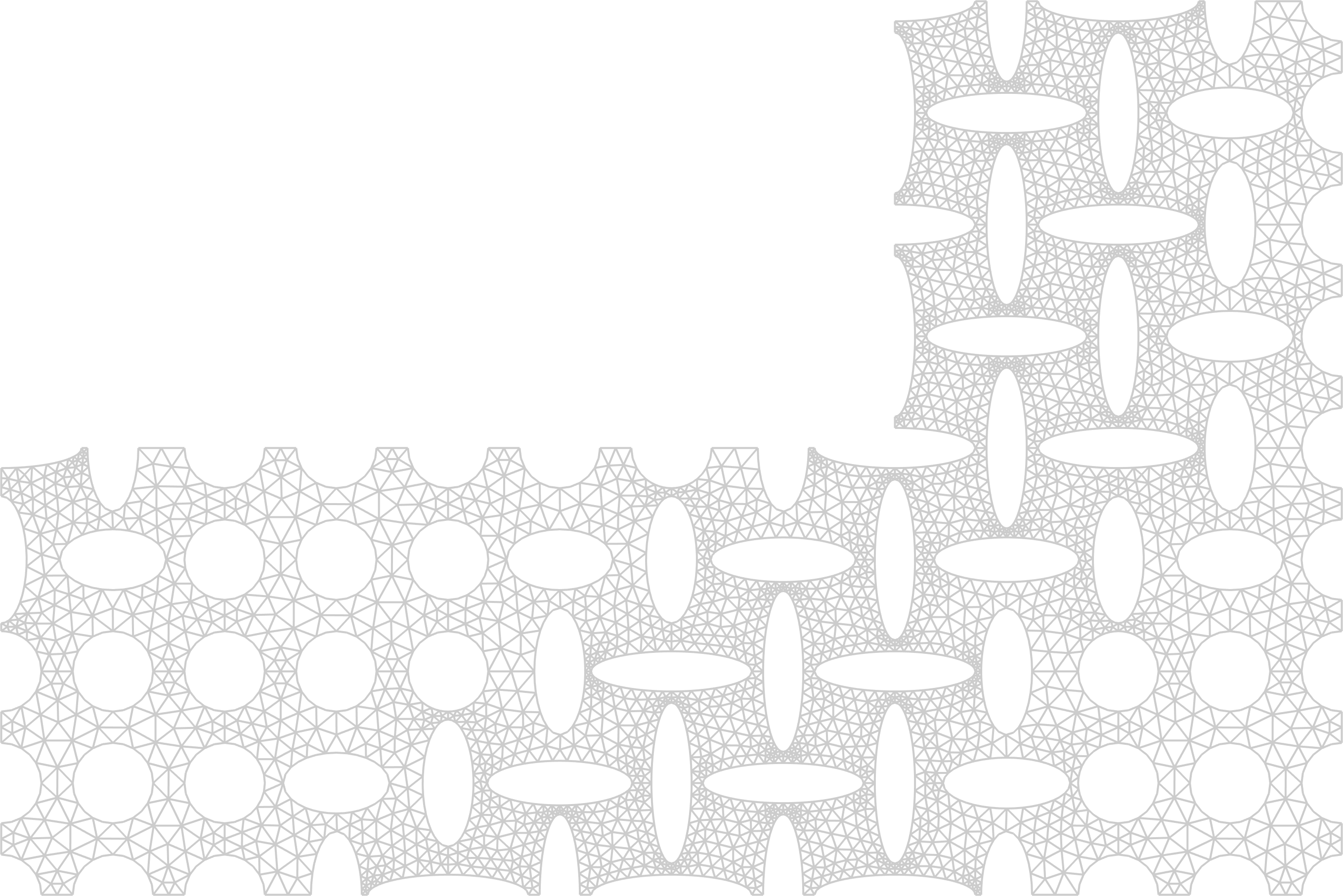} &
		\includegraphics[width=0.35\textwidth]{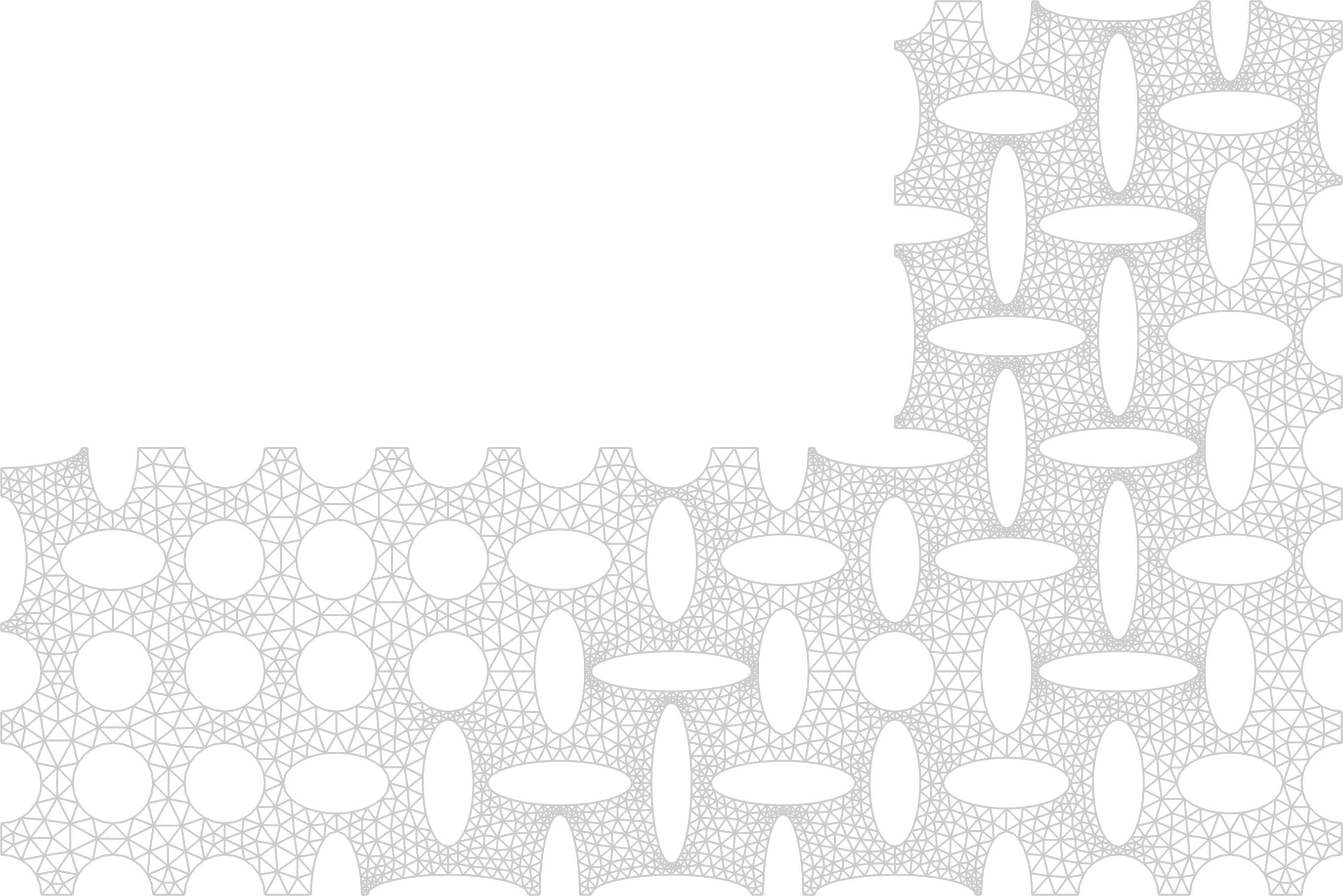} \\
		$\phi = -6.220^{\circ}$ & $\phi = -6.361^{\circ}$ \\
		(a) & (b)
	\end{tabular}
	\caption{\hl{Comparison of undeformed designs arising from the alternative factorial-search setups: (a) the same split as the original setup from the main manuscript with opposite initial vertex codes in the fixed region and (b) the setup with extended initial region prolonged to contain the whole bottom part of the domain. The first row lists designs attaining maximal tilt angle, the bottom row contains designs for the minimal tilt angle. Note that the final designs in column (a) are identical to the original designs from Fig.~2 from the main manuscript.}}
	\label{fig:different_factorial_setup}
\end{figure}

\hl{The final designs used in the main manuscript were pertinent to the particular setup during the factorial search. To investigate the effect of the setup on the resulting extremes, we ran two additional factorial searches with different setups.

In the first setup variation, we kept the split from Fig.~1c in the main manuscript but we considered the opposite initial vertex codes in the fixed region, i.e., we started with circular holes in region $\textsf{BCGFE}$ first. Even though the interim results were different ($\phi = -6.009^{\circ}$ vs. $\phi = -5.885^{\circ}$ for the design minimizing the tilt angle and $\phi = 6.405^{\circ}$ vs. $\phi = 6.601^{\circ}$ for the design maximizing the tilt), the final identified extreme designs were identical to those used for the assembly and experiments in the main manuscript.

For the second setup variation, we extended the region of the initial evaluation such that it included all vertex positions in the bottom part (the region marked by vertices $\textsf{A}$-$\textsf{B}$-$\textsf{C}$-middle of edge $\textsf{GC}$-$\textsf{E}$-$\textsf{D}$ in Fig.~1c), while keeping elliptical holes in the remaining six vertex positions. Again, this was then followed by altering the codes in the remaining part. As shown in~\mbox{\Fref{fig:different_factorial_setup}}, the final designs differed from the ones reported in the main manuscript; however, the differences in the tilt angles were small. The relative difference was below 0.5\textperthousand{} for the maximal tilt and 1.1\% for the minimal angle (relative to the original tilt span).}

\section{Fabrication}
The modularity that appears in the geometrical and numerical models encourages robot-assisted manufacturing. In our previous paper~\cite{nezerka_jigsaw_2018}, specimens were created directly from a single module equipped with jigsaw puzzle-like locks. However, to achieve an assembled system response that would be comparable to a monolithic assembly, the puzzle locks needed to be very tight, making them almost useless for robot-assisted assembly and disassembly. For this reason, here we included modularity in the manufacturing process one step earlier, i.e., we manufactured the final specimens by casting two-component Havel Composites ZA22 silicone rubber into automatically assembled molds.

\subsection{Molds}
In order to facilitate robot-assisted manufacturing, we devised modular molds that were composed of individual mold blocks. Since modularity reminded us of children's toys, we utilized the LEGO\textsuperscript{\textregistered} system---a standard $32 \times 32$ stud wide LEGO\textsuperscript{\textregistered} baseplate, in particular---as a way to attach individual blocks together. 
Such a setup allows the mold blocks to be reused for specimen molds with different shapes and internal compositions.

\begin{figure}
	\centering
	\begin{tabular}{cc}
		\includegraphics[width=0.35\textwidth]{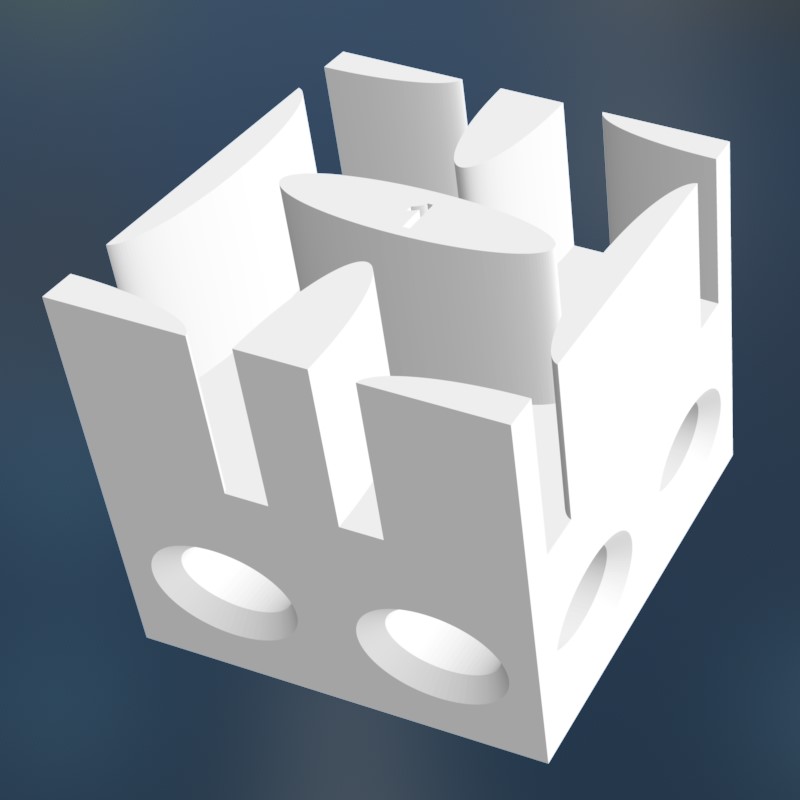} &
		\includegraphics[width=0.35\textwidth]{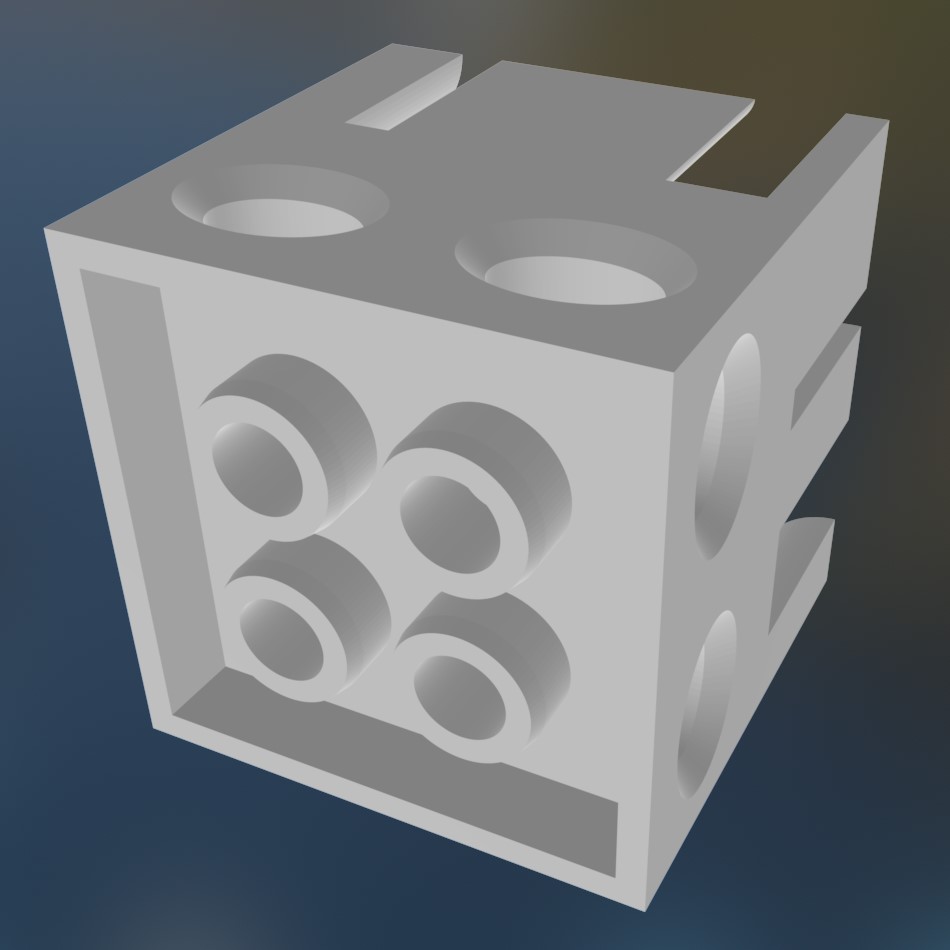}\\
		(a) & (b)
	\end{tabular}
	\caption{A 3D model of a mold block that corresponds to the bottom right module in Fig. 1b (in the main manuscript): (a) a top view at the part comprising parts of ellipses, (b) a bottom view at the part designed to attach to a LEGO baseplate.}
	\label{fig:mold_block}
\end{figure}

Following LEGO measurements, the dimensions of each mold block were 23.8 $\times$ 23.8~mm horizontally and 19.6~mm vertically. 
Each mold block consisted of three horizontally-stacked parts.
The top part was formed by ellipses (or their parts), which corresponded to the holes in material modules, see Fig. 3a in the main manuscript and \Fref{fig:mold_block}a. This part was automatically generated from the module geometries.
The middle part contained a pair of elliptical holes on each side, designed to facilitate robotic manipulation with a custom-made gripper, see Inset 2 in Fig. 3b in the main text.
Each hole lead through the whole block and was conically widened at both ends to compensate for spatial inaccuracies during manipulation.
Finally, the bottom part of each mold block followed from the standard design of a 3$\times$3 stud wide LEGO brick such that our mold block could be easily attached to a LEGO baseplate, see~\Fref{fig:mold_block}b.
Furthermore, we tapered each stud conically at the bottom of the mold block to make it easier for the robot to securely click the blocks in the right position.

In addition to the regular mold blocks described above, we also created three types of auxiliary blocks, which played the role of a formwork around the specimen boundaries to prevent leakage of uncured silicone rubber from the mold; see the black blocks in Figures 4a--4c in the main manuscript.
All blocks were produced from a PLA material with a conventional Prusa i3 MK3S+ hobby 3D printer. 

\subsection{Robot-assisted manufacturing}
\begin{figure}
	\centering 
	\scalebox{0.775}{%% Creator: Inkscape 1.2.1 (9c6d41e410, 2022-07-14), www.inkscape.org
%% Accompanies image file 'robot_composite.pdf' (pdf, eps, ps)
\begingroup%
  \makeatletter%
  \providecommand\color[2][]{%
    \errmessage{(Inkscape) Color is used for the text in Inkscape, but the package 'color.sty' is not loaded}%
    \renewcommand\color[2][]{}%
  }%
  \providecommand\transparent[1]{%
    \errmessage{(Inkscape) Transparency is used (non-zero) for the text in Inkscape, but the package 'transparent.sty' is not loaded}%
    \renewcommand\transparent[1]{}%
  }%
  \providecommand\rotatebox[2]{#2}%
  \newcommand*\fsize{\dimexpr\f@size pt\relax}%
  \newcommand*\lineheight[1]{\fontsize{\fsize}{#1\fsize}\selectfont}%
  \ifx\svgwidth\undefined%
    \setlength{\unitlength}{595.27559055bp}%
    \ifx\svgscale\undefined%
      \relax%
    \else%
      \setlength{\unitlength}{\unitlength * \real{\svgscale}}%
    \fi%
  \else%
    \setlength{\unitlength}{\svgwidth}%
  \fi%
  \global\let\svgwidth\undefined%
  \global\let\svgscale\undefined%
  \makeatother%
  \begin{picture}(1,0.45085467)%
    \lineheight{1}%
    \setlength\tabcolsep{0pt}%
    \put(0,0){\includegraphics[width=\unitlength,page=1]{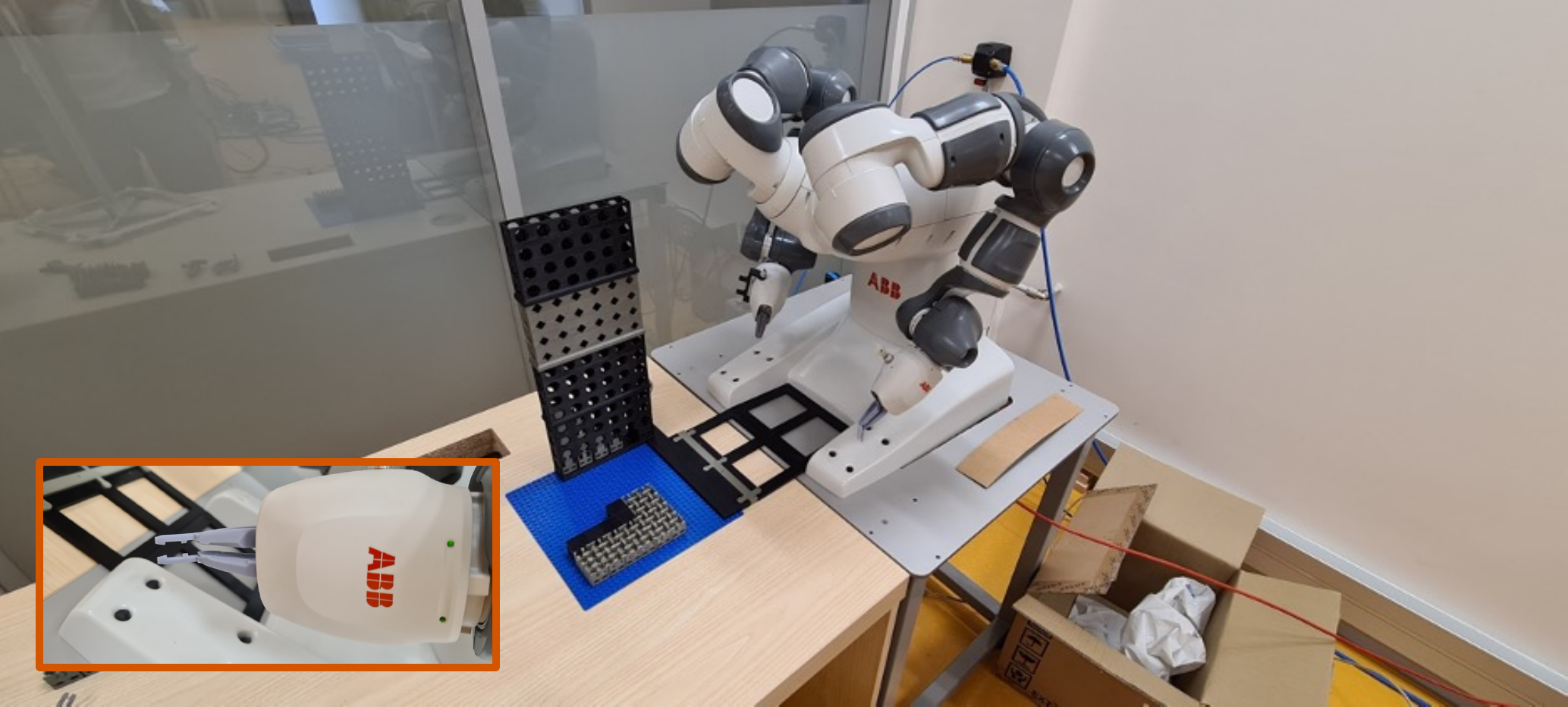}}%
    \put(0.02529312,0.16637054){\color[rgb]{0.83137255,0.33333333,0}\makebox(0,0)[lt]{\lineheight{1.25}\smash{\begin{tabular}[t]{l}Inset 1\end{tabular}}}}%
    \put(0,0){\includegraphics[width=\unitlength,page=2]{robot_composite.pdf}}%
    \put(0.88704878,0.14527766){\color[rgb]{0.83137255,0.33333333,0}\makebox(0,0)[lt]{\lineheight{1.25}\smash{\begin{tabular}[t]{l}Inset 2\end{tabular}}}}%
    \put(0,0){\includegraphics[width=\unitlength,page=3]{robot_composite.pdf}}%
    \put(0.88725542,0.38598819){\color[rgb]{0.83137255,0.33333333,0}\makebox(0,0)[lt]{\lineheight{1.25}\smash{\begin{tabular}[t]{l}Inset 3\end{tabular}}}}%
  \end{picture}%
\endgroup%
} 
	\caption{The complete robot-assisted assembly setup with details of (i) the pair of 3D-printed grippers (Inset 1), (ii) 3D model of the baseplate and silos with distancers (Inset 2), and (iii) detail of silos (Inset 3).}
	\label{fig:robot}
\end{figure}

For the automated assembly, we utilized an ABB IRB 14000 YuMi dual-arm collaborative robot, shown in~\Fref{fig:robot}.
As already mentioned in the previous section, the robot was equipped with the custom-made gripper mounted on a robotic arm to manipulate the blocks, similar to the motions of a pallet truck.
Since the mold block featured holes for the gripper on its sides, any block assembly could be achieved in scan-line order without colliding with already placed mold blocks.

The mold blocks were stored in 3D printed silos that were designed to facilitate easy access for the gripper.
Additionally, the silos were stackable, allowing for a potential increase in storage space, see Inset 3 in~\Fref{fig:robot}. Each silo bay stored multiple types of mold blocks, and the robot was instructed a priori regarding where to reach for blocks according to a given assembly plan. 
To provide the robot with a known and stable position reference to all components, 3D printed distancers were used to secure the position of the LEGO baseplate, which accommodated both the silos and blocks, see Inset 2 in~\Fref{fig:robot}.

The robot was programmed in its native RAPID programming language using the RobotStudio software. The resources including source codes and 3D models are provided in a Zenodo repository~\cite{doskar_2023_zenodo}.

Thanks to the distancers mentioned above, only two robot arm positions (variables \texttt{silo\_point}, \texttt{placement\_position}) needed to be set. This was done manually by placing the robot arm into the desired positions and saving the coordinates. However, in order to avoid singular joint positions of the YuMi arm, we set two additional positions: \texttt{home\_point}, which was the starting and ending position of the arm, and \texttt{base\_point}, which stored a position above the build plate used to get to and from the \texttt{home\_point}.

The main structure of the program was as follows
\begin{verbatim}
	PROC main()
	goHome;
	goBase;
	FOR i FROM 1 TO Dim(cubePlacement,1) DO
	get_cube i;
	place_cube i;
	ENDFOR
	goBase;
	goHome;
	ENDPROC
\end{verbatim}
where the \texttt{cubePlacement} input variable was a numeric matrix that stored integers \texttt{x\_id}, \texttt{y\_id}, and \texttt{silo\_id} of one mold block to be placed in each row. Integers \texttt{x\_id} and \texttt{y\_id} referred to the position on the baseplate with respect to the \texttt{placement\_position}, shown in~Inset 1 in~\Fref{fig:robot} as a green tile with coordinate axes. 
The physical setup limits were \texttt{x\_id} $\in \interval{0}{8}$, \texttt{y\_id} $\in \interval{0}{5}$ and \texttt{silo\_id} $\in \interval{1}{6}$. Note the construction area was offset from the silos in the Y direction to define the robot's handling space; therefore, the span of \texttt{y\_id} was smaller than \texttt{x\_id}.

\section{Experimental validation}

\subsection{Material}
\label{sec:validation_mat}

\begin{figure}
	\centering
	\includegraphics[width=0.95\textwidth]{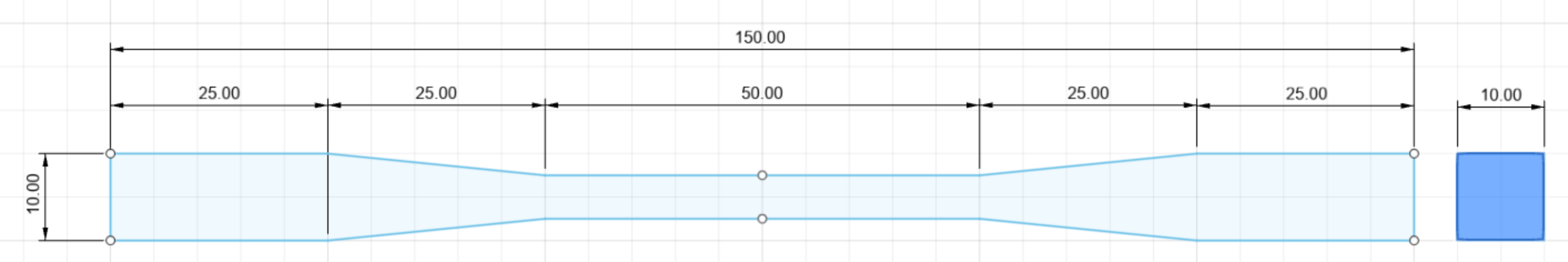} 
	\caption{Dimensions of a dogbone specimen (in millimeters) used for the uniaxial tensile test.}
	\label{fig:dogbone_dimensions}
\end{figure}

For initial design exploration, the material parameters for the hyper-elastic law (1) were set to $c_{1} = 0.55~\text{MPa}$, $c_{2} = 0.3~\text{MPa}$, and $K = 55~\text{MPa}$ as reported by Bertoldi et al.~\cite{bertoldi_mechanics_2008} and used, e.g., in~\cite{rokos_extended_2020,van_bree_newton_2020}.
However, driven by discrepancies between preliminary experiments and numerical models, we performed a series of uni-axial tensile tests on dogbone silicone specimens cast from Havel Composites ZA22, the material used to produce the L-shape specimens. Length of the dogbone specimen was 150 mm, the end parts were 10 mm wide and tall, while the middle cross-sectional dimensions were 5$\times$5 mm; see \Fref{fig:dogbone_dimensions} for details.
The MTS Criterion loading frame model 43 equipped with a load cell of maximum capacity 500~N was used for the tensile tests. 
We measured displacements using Digital Image Correlation (DIC) method during the load test. Due to the low-adhesion surface of the silicone sample, we could not use standard spray paints. Eventually, we applied a special powder paint (HELLING Standard-Check developer Medium Nr.~3) directly to the silicone surface as a white base layer. After complete drying, it was then possible to apply randomly scattered speckles using a standard black spray paint, see~\Fref{fig:machine_setup}.
We tested three specimens made out of the same batch of silicone polymer used to produce the L-shape samples.

\begin{figure}
	\centering
	\includegraphics[width=0.45\textwidth]{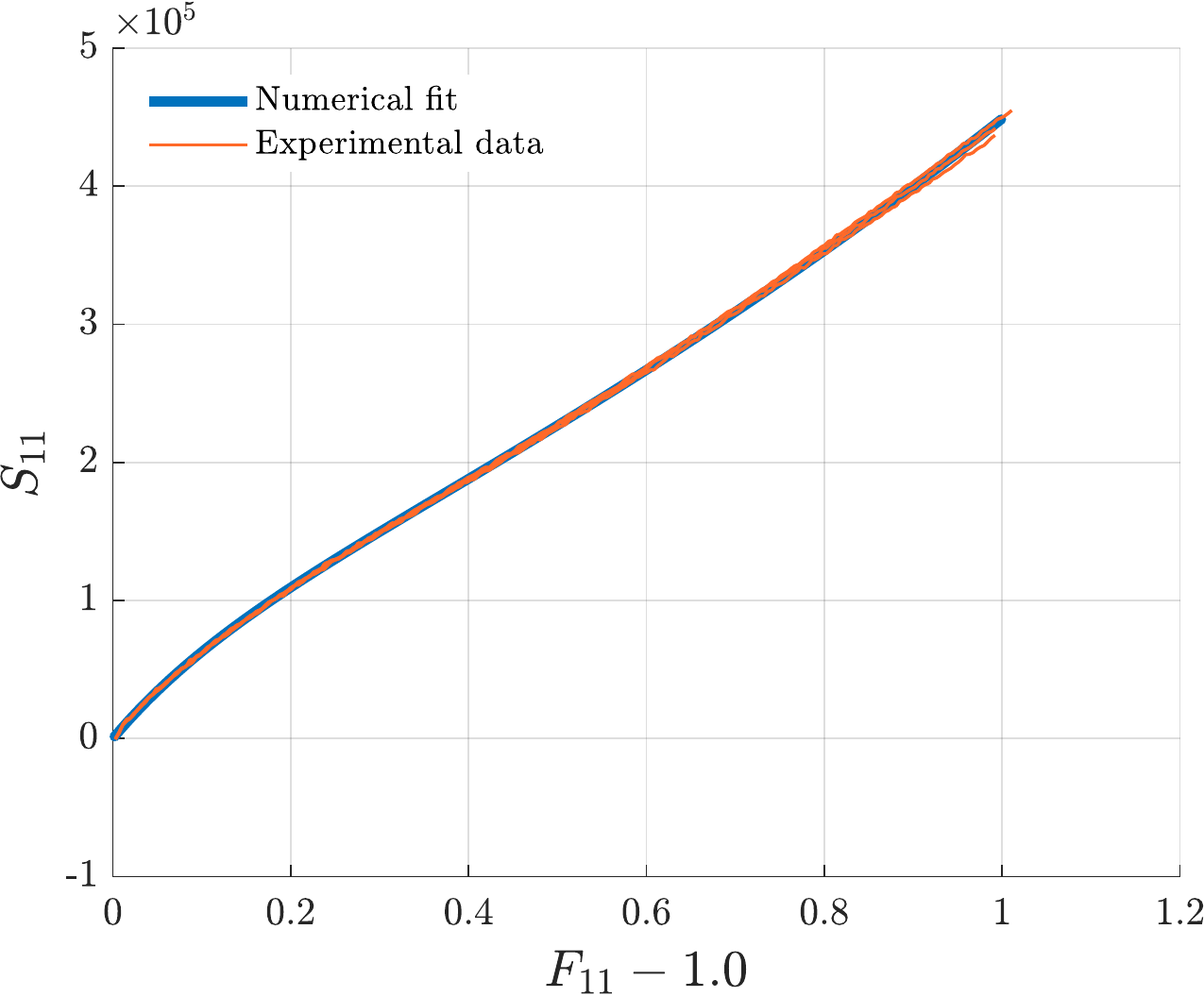}
	\caption{Comparison of stress-strain curves expressed in terms of $11$-components of deformation gradient $\tenss{F}$ and second Piola-Kirchhoff stress $\tenss{S}$ obtained from (i) the experimental uniaxial tensile tests (thin red lines) and (ii) numerical model with optimized material parameters (in blue). }
	\label{fig:parameter_fitting}
\end{figure}

\begin{figure}
	\centering
	\begin{tabular}{cc}
		\includegraphics[width=0.3\textwidth]{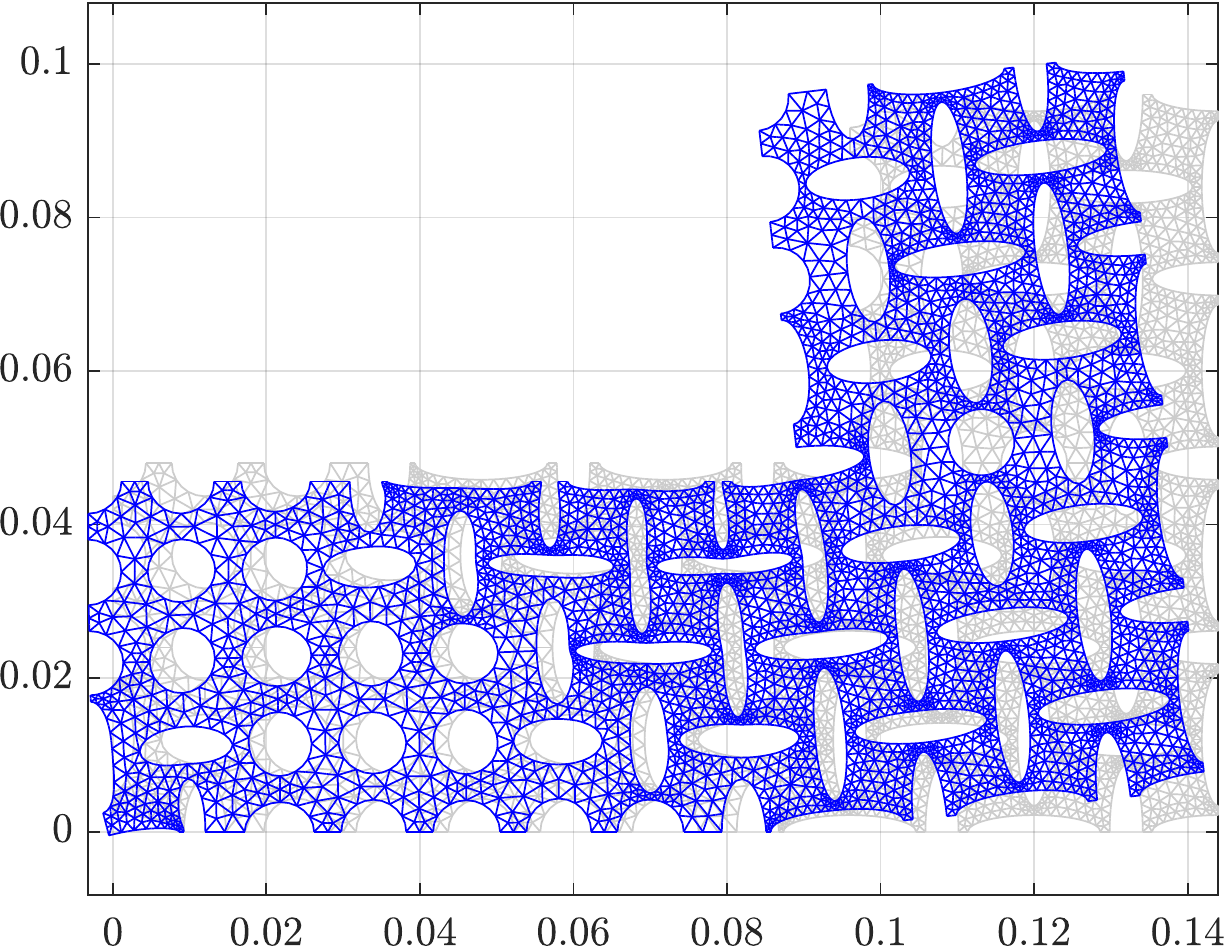} & 
		\includegraphics[width=0.3\textwidth]{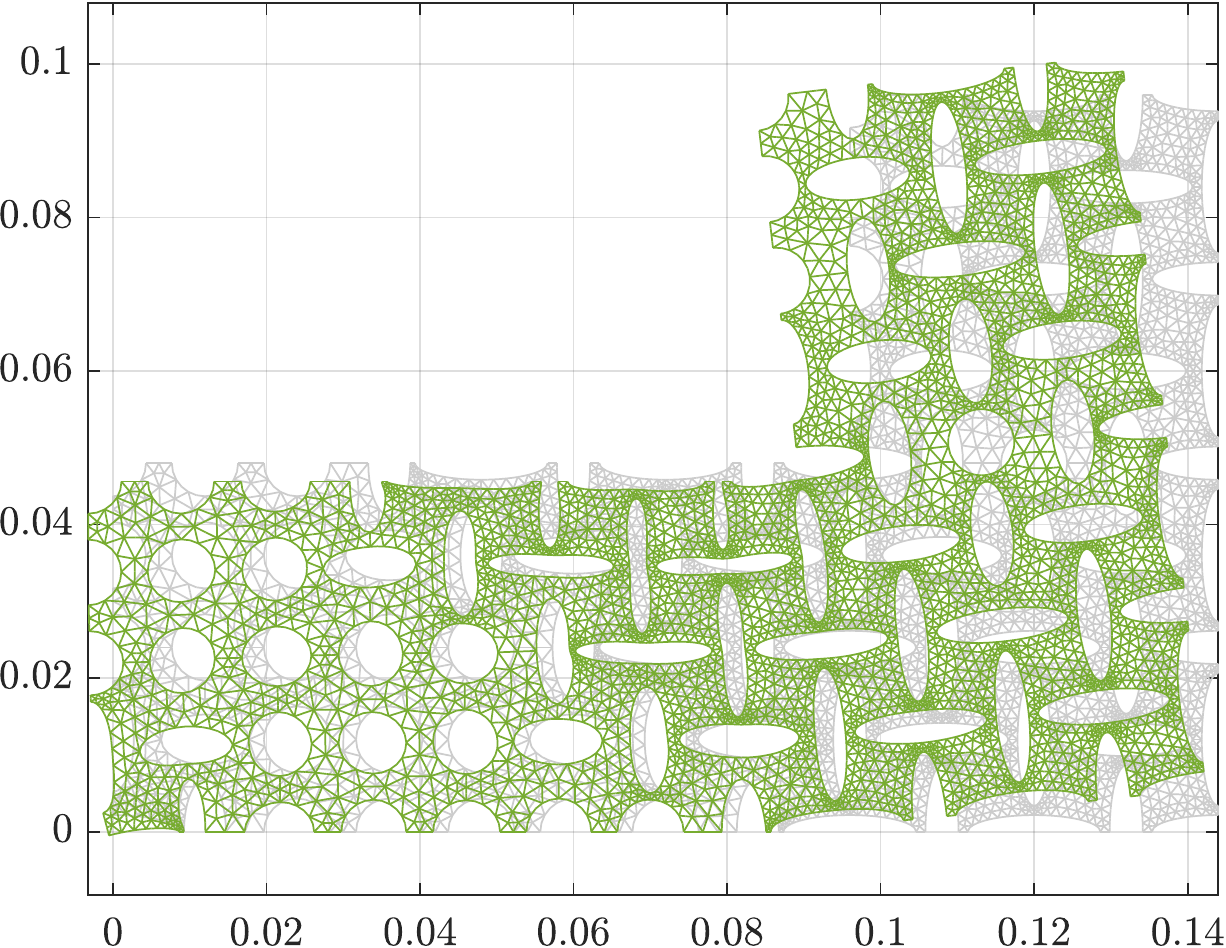} \\
		\includegraphics[width=0.3\textwidth]{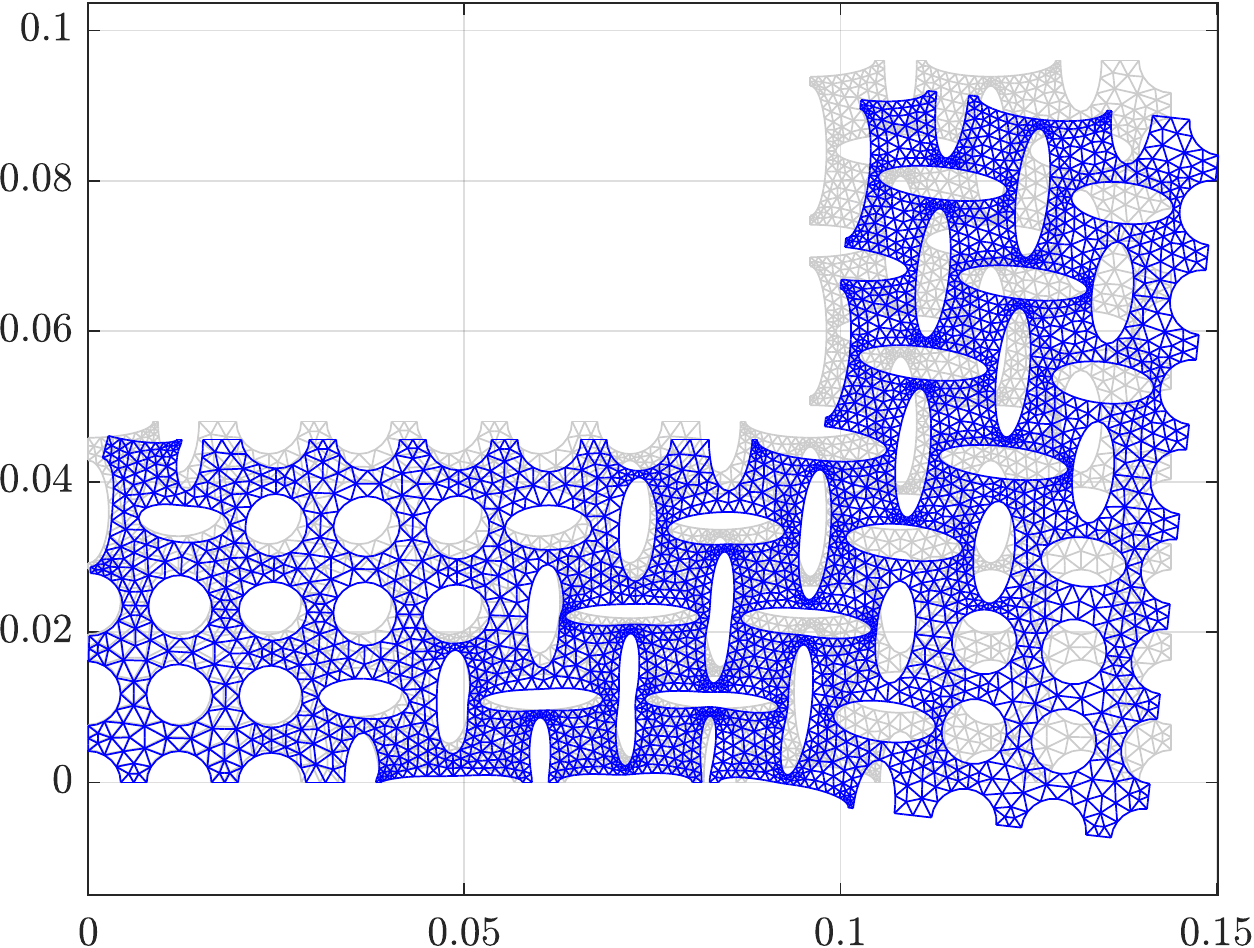} & 
		\includegraphics[width=0.3\textwidth]{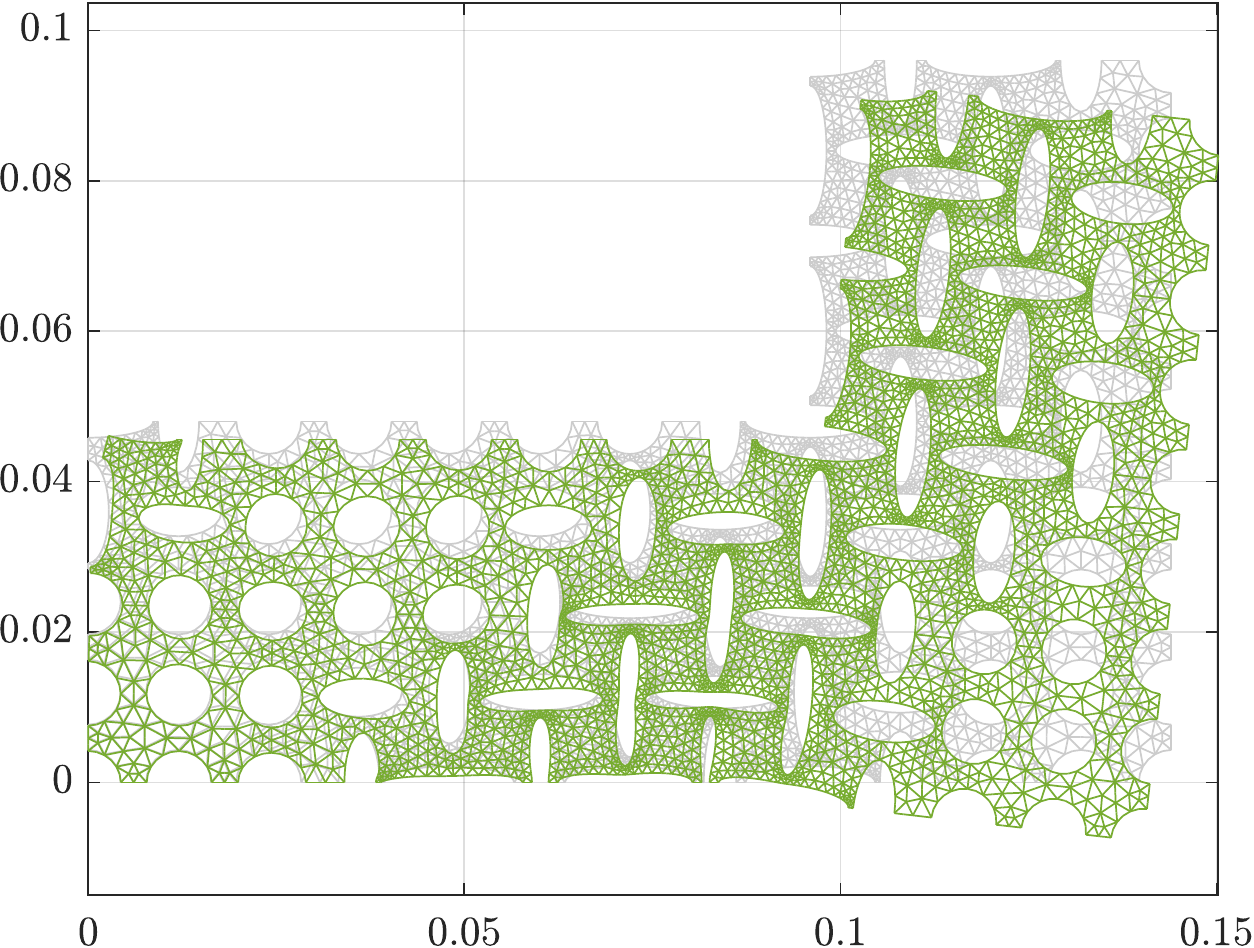} \\
	\end{tabular}
	\caption{Comparison of numerical predictions for the original material parameters (left) and with the optimized material coefficients (right), both for modular layouts maximizing (top row) and minimizing (bottom row) the tilt angle $\phi$.}
	\label{fig:influence_of_parameters}
\end{figure}

\begin{figure}
	\centering
	\begin{tabular}{cc}
		\includegraphics[height=0.35\textwidth]{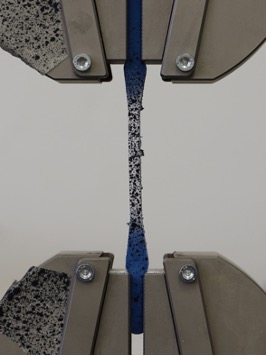} 
		&
		\includegraphics[height=0.35\textwidth]{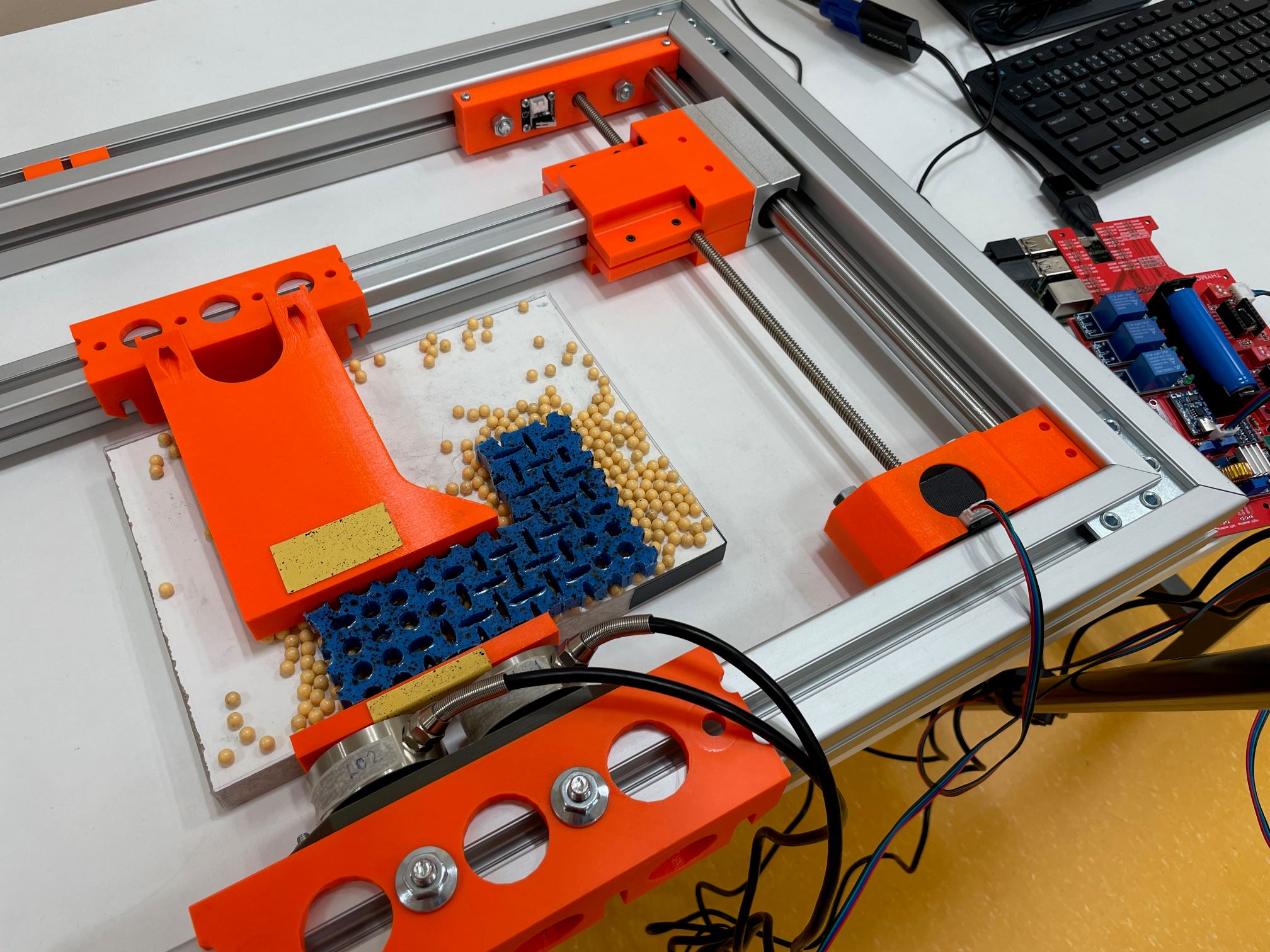}\\
		(a) & (b) 
	\end{tabular}
	\caption{Details of (a) the contrast pattern on the dogbone specimen's surface during the uniaxial tensile test and (b) the loading part of the Thymos test rig with the contrast pattern applied to an L-shape specimen (the overall image of the Thymos loading machine is shown in Fig.~4d in the main manuscript).}
	\label{fig:machine_setup}
\end{figure}

The obtained stress-strain curves, shown in red in~\Fref{fig:parameter_fitting}, were subsequently fitted with the material model of form (1), yielding the final coefficients $c_{1} = 0.0977~\text{MPa}$, $c_{2} = 0.0315~\text{MPa}$, and $K = 9.77~\text{MPa}$.
However, since the ratio between coefficients remained unaltered, the change in material parameters had negligible impact on simulation results; compare the deformed shapes of both extreme assemblies with the original and optimized material parameters in \Fref{fig:influence_of_parameters}. The reason for this is that the loading scenario is displacement-driven and the quantities of interest (the tilt angle $\phi$, in particular) are also geometrical.

\subsection{L-shape specimen}

\hl{To mitigate the influence of self-weight on in-plane deflection and to suppress out-of-plane displacement caused by inaccurate in-plane alignment of the sample in the loading machine, the L-shape was tested in a horizontal setup. To this end, }
we developed our own horizontal, Thymos\footnote{\href{http://thymos.cz/}{thymos.cz}} open-hardware testing rig comprising a frame of aluminum bars connected by 3D printed fasteners and screws. It was fitted with a pair of load cells with a maximum capacity of 50~N. The loading frame used custom-made 3D printed load plates to which the sample was attached. Loading was facilitated via a sliding crossbar driven by a pair of stepper motors, a threaded rod, and rails. The whole setup was controlled with a combination of an Arduino microcontroller and a single-board Raspberry Pi computer.

Again, DIC was used to obtain full-field displacement data for the specimen in all phases of testing.
In particular, we used a Canon EOS 6D Mark ii digital DSLR camera with a Canon EF 24-70 mm f/4 L IS USM lens to capture images with 6240$\times$4160~px resolution. DIC calculations were performed in Ncorr, an open source 2D DIC MATLAB program~\cite{Blaber_2015_Ncorr}, and the results were post-processed and evaluated using the Ncorr\textunderscore post tool~\cite{Nezerka_2016_IEN}.

As already mentioned in~\ref{sec:validation_mat}, an initial comparison revealed a discrepancy between numerics and experiments, which was not caused by the originally stiffer material parameters used in simulations. Eventually, friction between the specimen and the test rig (including its base) was identified as the main mechanism causing the discrepancy.
As a remedy, we placed the specimens on loosely placed plastic airsoft pellets (6-mm diameter) playing the role of planar ball bearings.
However, due to construction limitations, the same adjustment could not be easily done for the machine's (vertical) loading and supporting plates, and thus we still observed stick-slip behavior of material ligaments in contact with these plates.
Incorporating a standard Coulomb friction model was not sufficient, as it necessitated significantly different friction coefficients for different contact areas to match the experiments. Eventually, we extracted a lateral displacement profile for each center point of the contact areas (highlighted with red squares in Fig.~5a in the main manuscript for one optimized assembly) from DIC and used these values as additional Dirichlet boundary conditions in the simulations.

\begin{figure}[!b]
	\centering
	\begin{tabular}{cc}
		\includegraphics[width=0.3\textwidth]{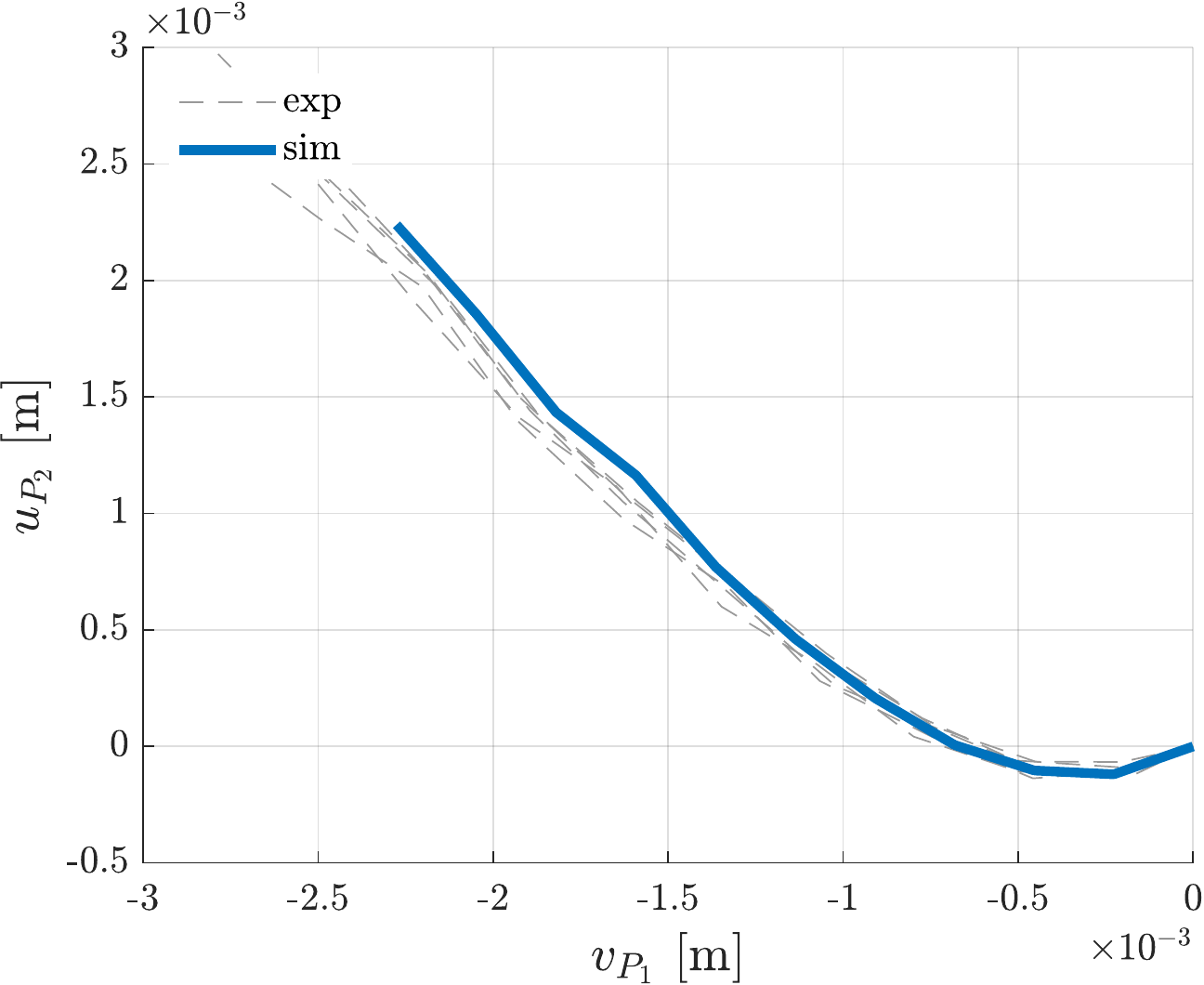} &
		\includegraphics[width=0.3\textwidth]{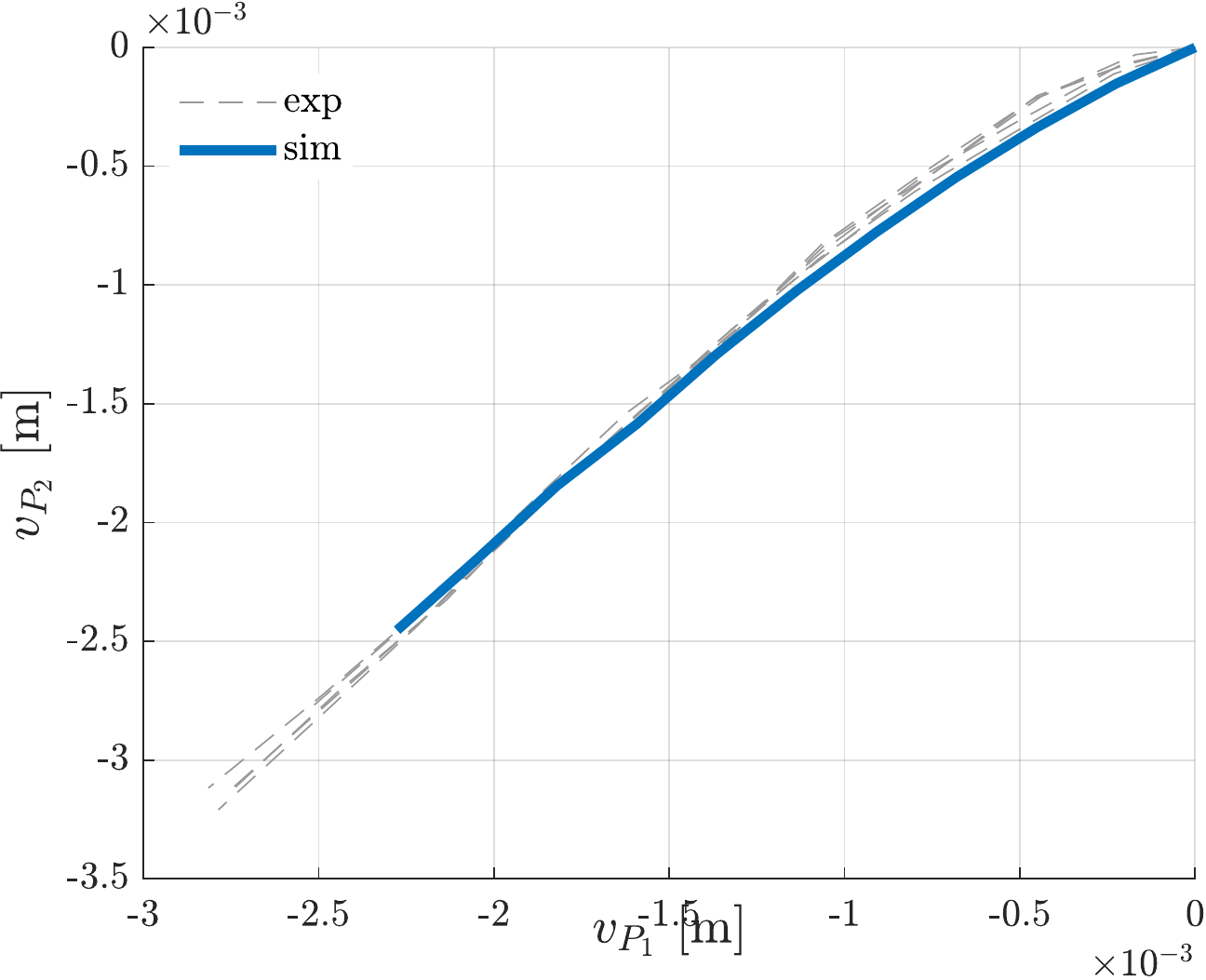} \\
		%(a) & (b) \\
		\includegraphics[width=0.3\textwidth]{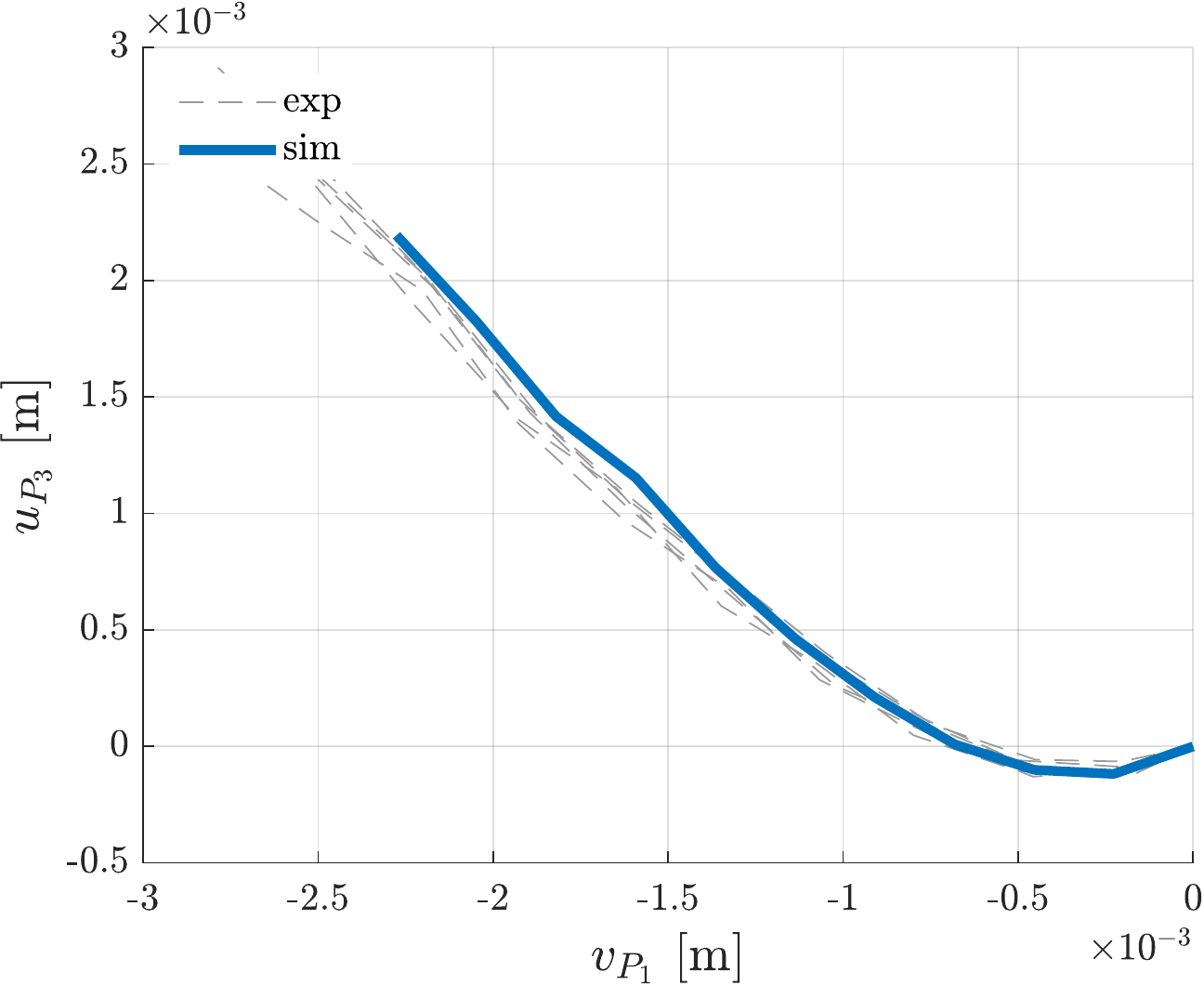} &
		\includegraphics[width=0.3\textwidth]{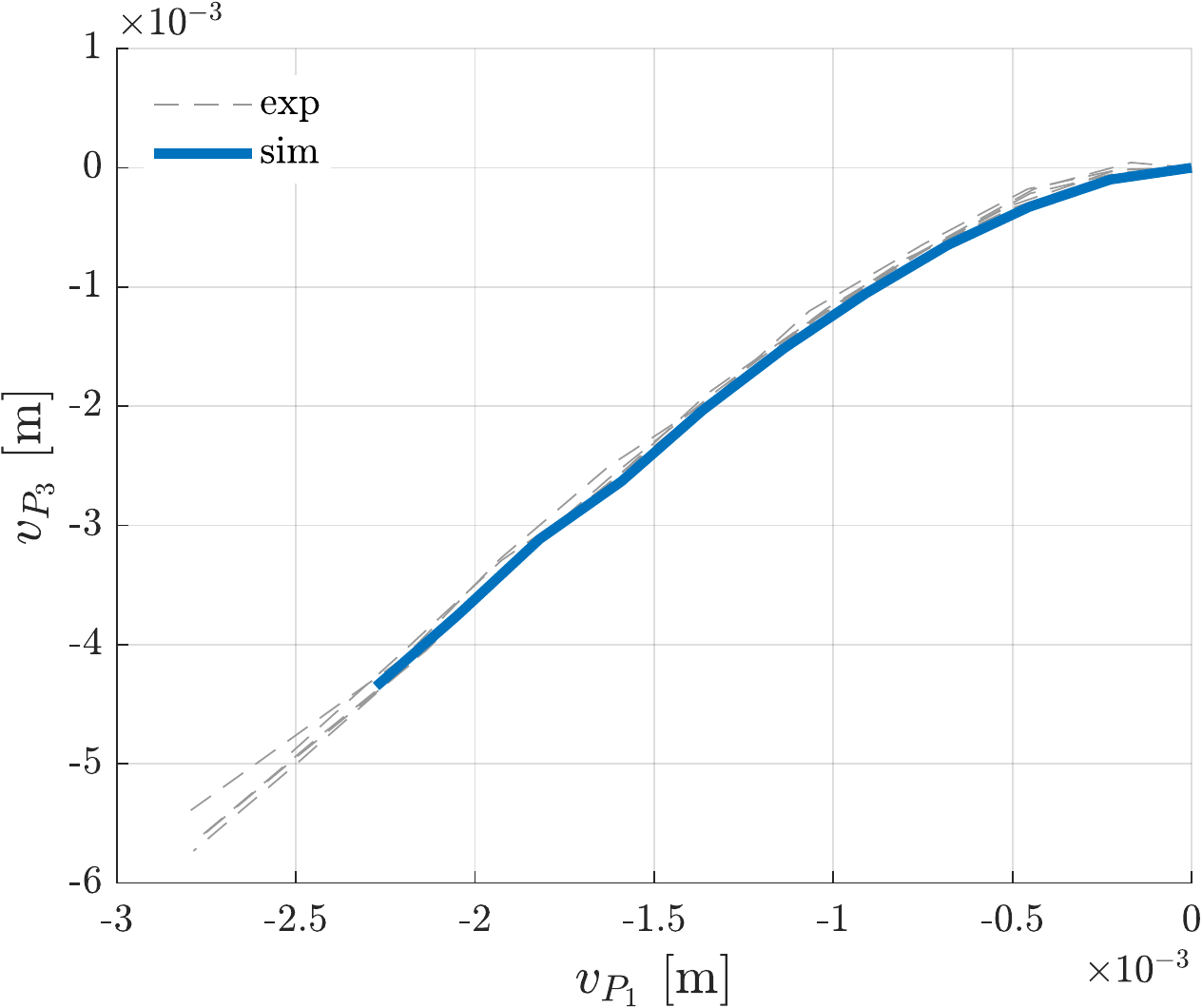} \\
		%(c) & (d) \\
	\end{tabular}
	\caption{Comparison of computed (in blue) against experimentally obtained (in dashed grey) displacements of points 2 and 3 from Figure 5 (in the main manuscript), parameterized by the prescribed vertical displacement of point 1. Data are shown for the case of modular layout minimizing the tilt angle $\phi$.}
	\label{fig:comparison_min}
\end{figure}

\begin{figure}[!t]
	\centering
	\begin{tabular}{cc}
		\includegraphics[width=0.3\textwidth]{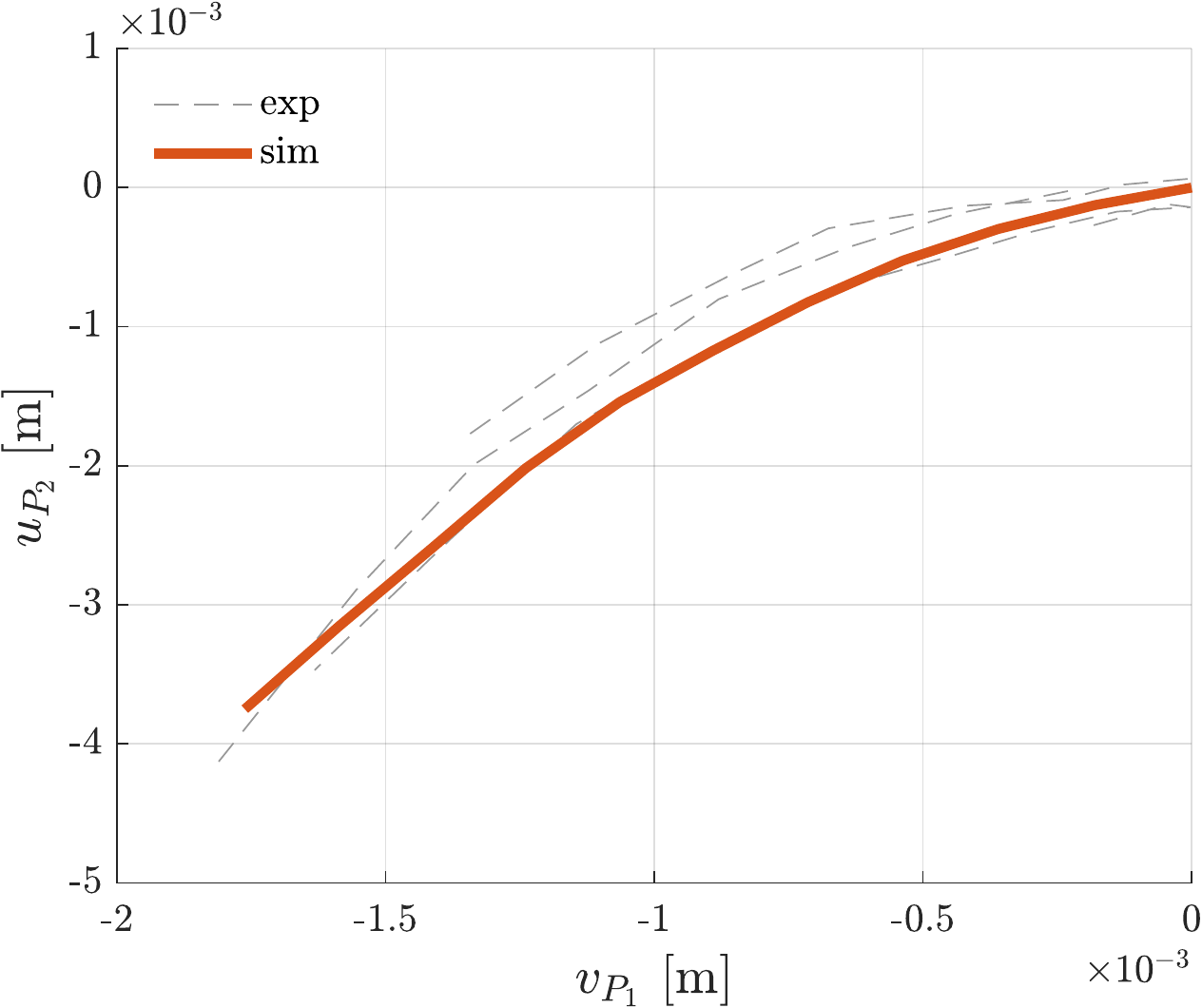} &
		\includegraphics[width=0.3\textwidth]{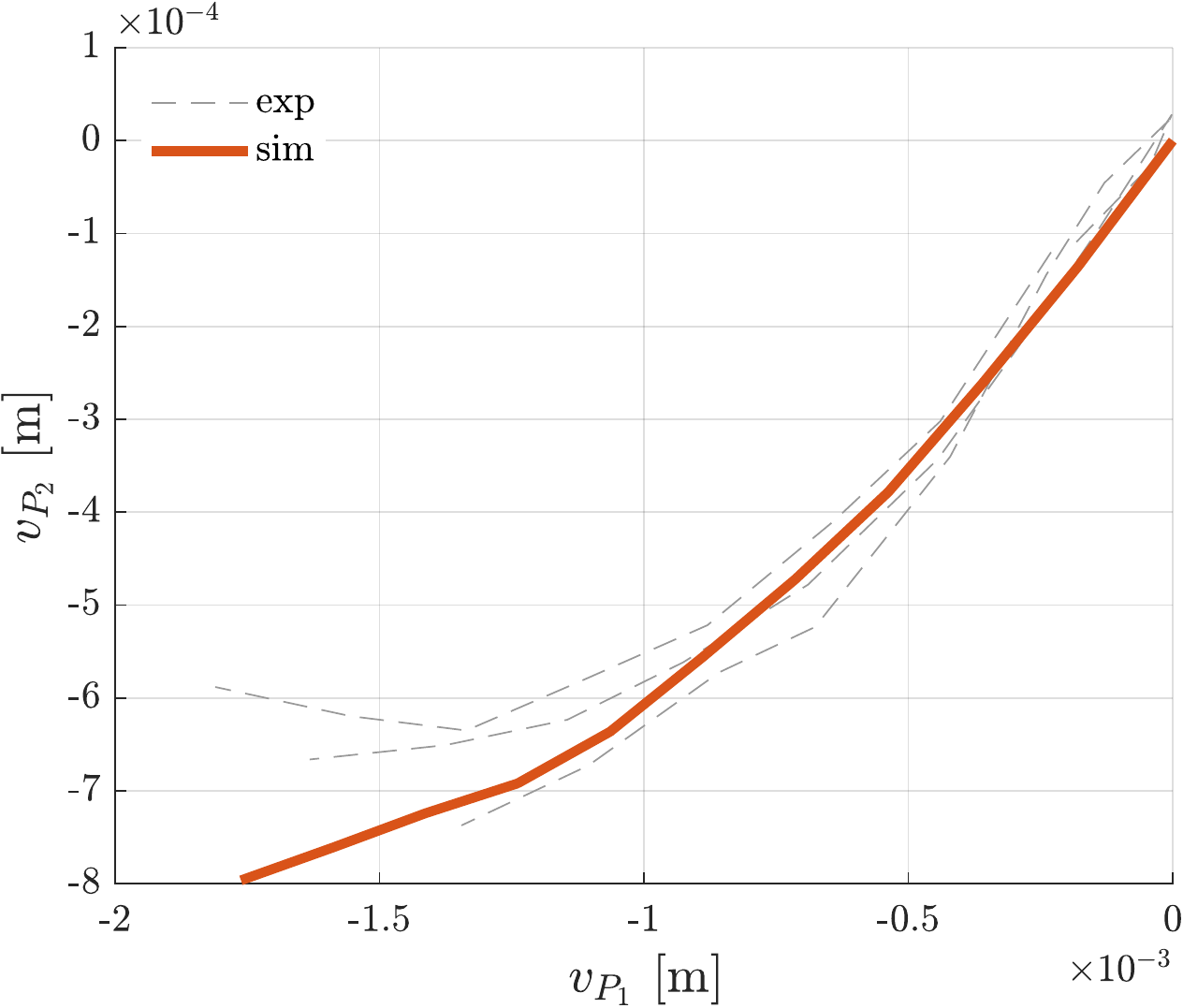} \\
		%(a) & (b) \\
		\includegraphics[width=0.3\textwidth]{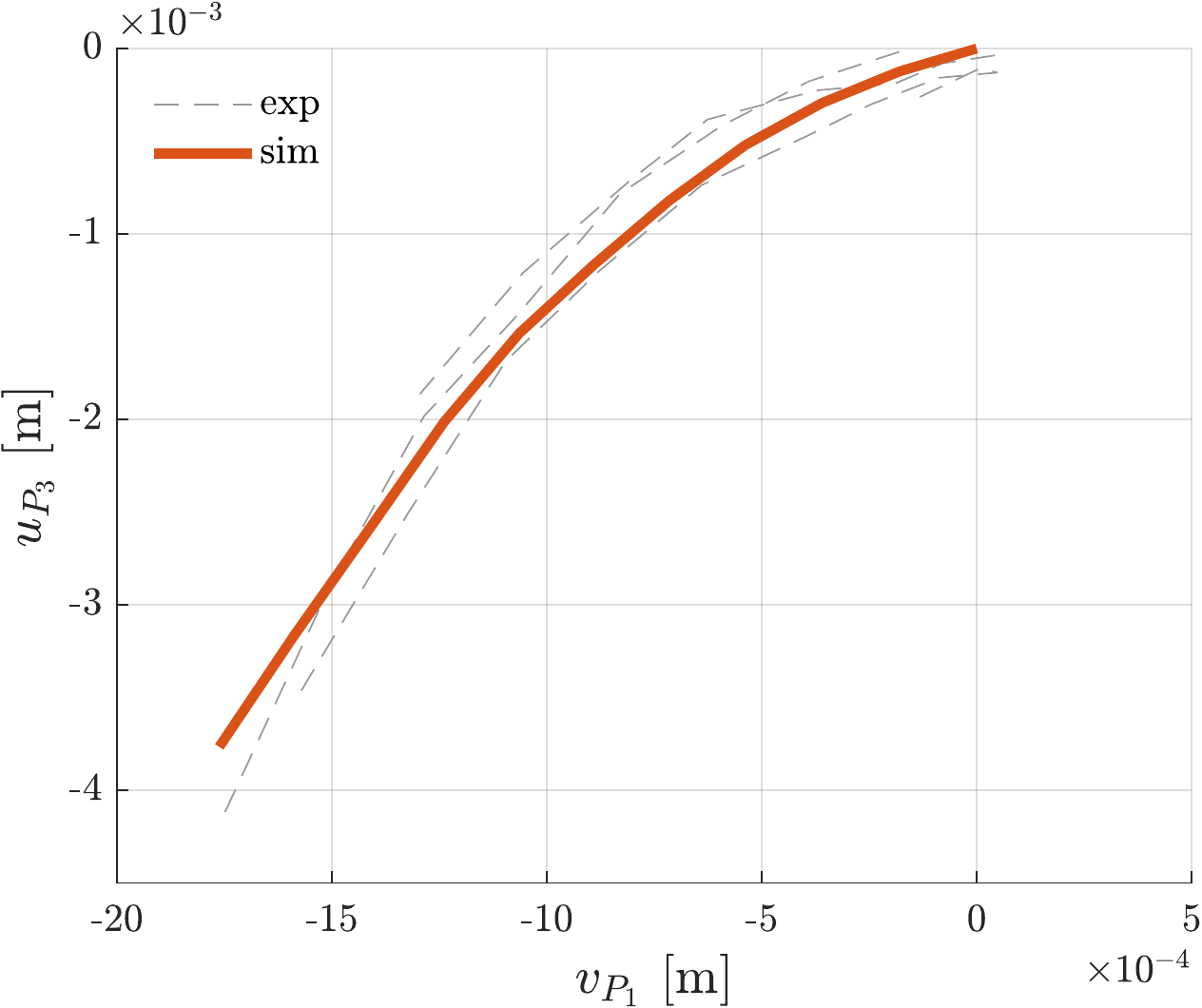} &
		\includegraphics[width=0.3\textwidth]{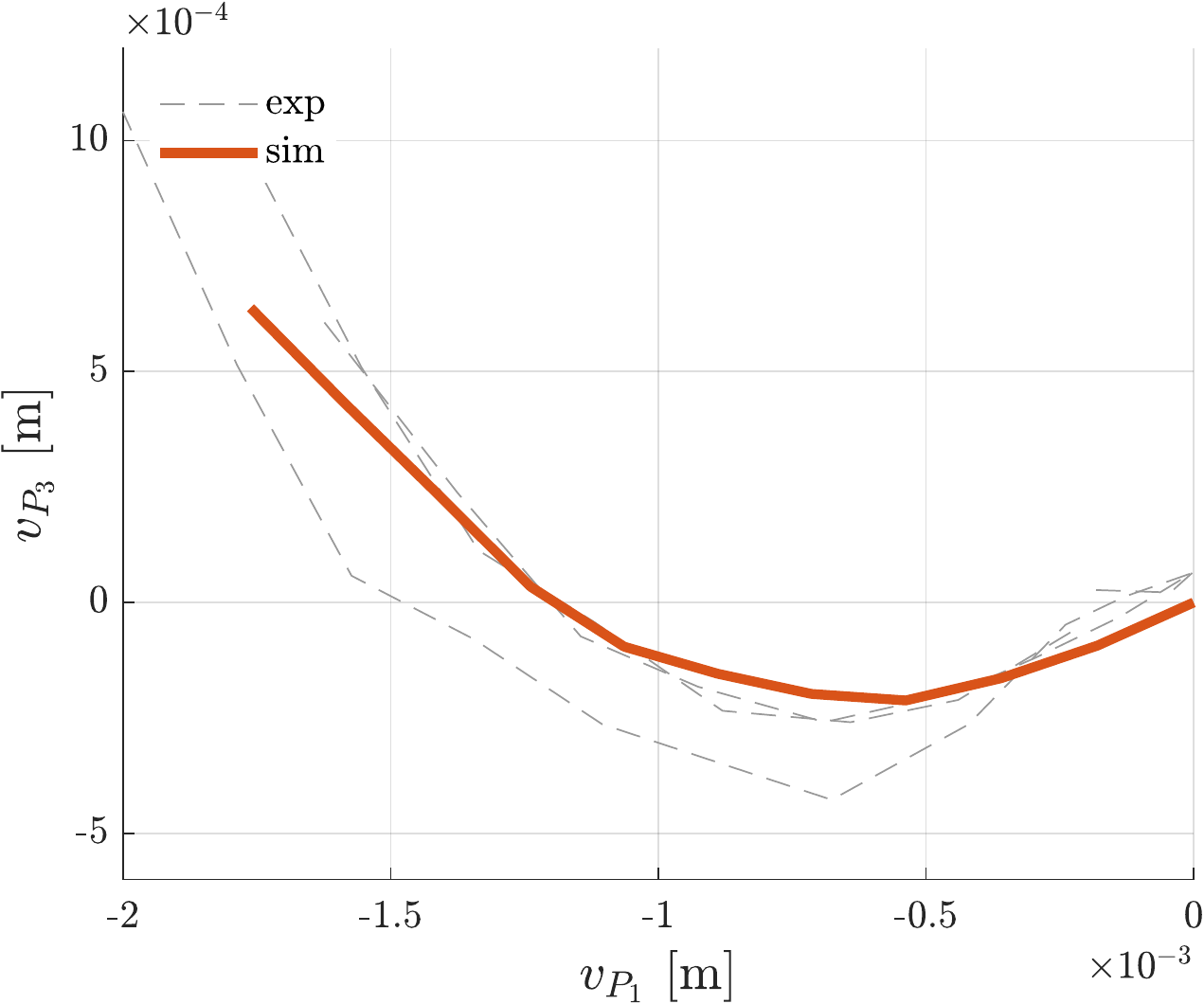} \\
		%(c) & (d) \\
	\end{tabular}
	\caption{Comparison of computed (in red) against experimentally obtained (in dashed grey) displacements of points 2 and 3 from Figure 5 (in the main manuscript), parameterized by the prescribed vertical displacement of point 1. Data are shown for the case of modular layout maximizing the tilt angle $\phi$.}
	\label{fig:comparison_max}
\end{figure}

Figures \ref{fig:comparison_min} and \ref{fig:comparison_max} provide a comparison of predicted and experimentally obtained vertical and horizontal displacements of two selected points $P_{2}$ and $P_{3}$, depicted in Fig.~5a (in the main manuscript), for the assembly minimizing (\Fref{fig:comparison_min}) and maximizing (\Fref{fig:comparison_max}) the tilt angle $\phi$. Experiments were repeated five and three times, respectively.
Both comparisons showed a very good agreement between the calculations and experiments with almost excellent match for the design minimizing the tilt angle and slightly higher scatter in experimental results in the case maximizing the tilt angle.

\begin{figure}
	\centering
	\includegraphics[width=9cm]{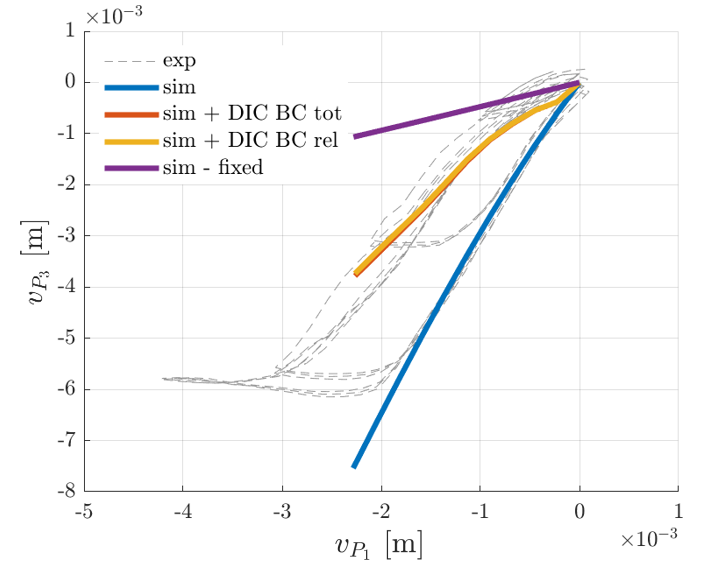}
	\caption{{Comparison of numerical simulations and experiments for cyclic loading of the design shown in Fig.~5 in the main manuscript (displacement labels are identical to those in~Fig.~5a). The figure contain: (i) results from a plain FE model without prescribed horizontal displacements (labeled as \enquote{sim}), (ii) results for two variants of simulations with prescribed horizontal displacements obtained from DIC (while the \enquote{sim + DIC BC tot} variant used the DIC-measured displacement directly, the \enquote{sim + DIC BC rel} variant corrected the displacement obtained with control measurements of known distances to compensate for settling of the loading plates), (iii) and results from a model with fixed horizontal displacement at the loading plates (labeled \enquote{sim - fixed}).}}
	\label{fig:cyclic_experiments}
\end{figure}

\hl{Note that the applied vertical displacement, in both simulations and experiments, closely approached the threshold at which self-contact initiated in both designs. Although the initial contact appeared between the ligaments along the loading plates---which can be imposed to the model via DIC measurements---the material's perforations also began closing shortly thereafter.
Experimental curves, akin to those in Fig.~5b of the main manuscript, from a cyclic loading combined with increase in the prescribed displacement are displayed in Figure~\ref{fig:cyclic_experiments}. The plateau at the highest loading level stems from the self-contact and the closure of the holes. 
Additionally, this figure highlights the impact of prescribed horizontal displacements acquired from DIC for intermediate load levels. Imposing the horizontal displacement at the contact ligaments led to a 50\% reduction in the vertical displacement at the upper section. Note also that the plain FE simulation reasonably aligned with the unloading branches. 
To complete this investigation, we also simulated response of the sample with all horizontal displacements along the loading plates fixed, the results of which are plotted in violet in Fig.~\ref{fig:cyclic_experiments}.}

% ----------------------------------------
% Bibliography
% ----------------------------------------